\documentclass[10pt,manuscript]{aastex}

\usepackage{multirow}
\usepackage{subfigure}
\usepackage[rightFloats]{fltpage}

\shorttitle{VARIABILITY FROM THE {\sl FIRST} SURVEY}
\shortauthors{Thyagarajan et~al.}

\begin{document}

\title{VARIABLE AND TRANSIENT RADIO SOURCES IN THE {\sl FIRST} SURVEY}


\author{NITHYANANDAN THYAGARAJAN\altaffilmark{1}, 
  DAVID J. HELFAND\altaffilmark{1}, 
  RICHARD L. WHITE\altaffilmark{2}
  AND ROBERT H. BECKER\altaffilmark{3,4}}

\altaffiltext{1}{Department of Astronomy, Columbia University, 
  Mail Code 5246, Pupin Hall, 550 West 120th Street, 
  New York City, NY 10027.}
\altaffiltext{2}{Space Telescope Science Institute, 
  3700 San Martin Drive, Baltimore, MD 21218.}
\altaffiltext{3}{Department of Physics, University of California, 
  Davis, 1 Shields Avenue, Davis, CA 95616-8677.}
\altaffiltext{4}{IGPP/Lawrence Livermore National Laboratory}
\email{t\_nithyanandan@astro.columbia.edu}


\begin{abstract}
A comprehensive search for variable and transient radio sources has been conducted using $\sim$~55,000 snapshot images of the {\sl FIRST} survey. We present an analysis leading to the discovery of 1,627 variable and transient objects down to mJy levels over a wide range of timescales (few minutes to years). Variations observed range from 20\% to a factor of 25. Multi-wavelength matching for counterparts reveals the diverse classes of objects exhibiting variability, ranging from nearby stars and pulsars to galaxies and distant quasars. Interestingly, more than half of the objects in the sample have either no classified counterparts or no corresponding sources at any other wavelength and require multi-wavelength follow-up observations. We discuss these classes of variables and speculate on the identity of objects that lack multi-wavelength counterparts.
\end{abstract}




\section{Introduction}\label{intro}

The first celestial radio source detected -- the Sun -- was discovered as a consequence of its variability \citep{hey46}. Nonetheless, while variability has continued to be a source of discovery and insight in the radio regime, few dedicated searches for variables and transients have been undertaken, primarily because the observing times required to conduct deep wide-field surveys are large and current instruments tend to have poor figures of merit for variability studies \citep{cor04}.

Indeed, one of the key scientific aims of the next generation of radio telescopes such as LOFAR, ASKAP, MWA and, eventually, the SKA is a study of radio variability. As discussed by \citet{hes09}, LOFAR will survey the sky for pulsars and fast transients at low frequencies (30-240~MHz) with an emphasis on timescales of less than a second. CVs, X-ray binaries, GRBs, SNe, AGN, flare stars, exoplanets and many more new phenomena, as well as extrinsic causes such as interstellar and interplanetary scintillation, are expected to contribute to its inventory of radio transients and variables \citep{fen08}. The Australian SKA Pathfinder \citep[ASKAP;][]{joh07,joh08} is designed to achieve very high survey speeds and noise levels down to $\sim 10-100$~$\mu$Jy in an hour in the 1~GHz band, and will shed light on GRBs, radio supernovae (RSNe), Intra-Day Variables (IDVs), etc. The Murchison Widefield Array (MWA) will operate in the 80-300~MHz range carrying out a blind search for variables and transients in addition to targeting explosive events, stellar and planetary phenomena, and compact objects over a range of timescales from nanoseconds to years \citep{lon09}. The EVLA \citep{nap06,rup00} will offer vastly improved sensitivity along with dynamic scheduling, providing a host of new capabilities for transient and variable searches. Finally, a search for radio variables and transients also forms an important part of the scientific objectives of the Square Kilometer Array (SKA). The SKA is expected to discover a number of classes of variable radio objects such as pulsars and magnetars, GRBs that are $\gamma$-ray loud and $\gamma$-ray quiet in both afterglow and prompt emission, sub-stellar objects such as brown dwarfs and exoplanets, microquasars, and potentially new classes of astrophysical phenomena \citep{cor04,cor08,wil04}. 

Most of the research on radio variability to date has focused on bright radio samples (e.g., \citet{gre86}, $S>0.4$~Jy at 408~MHz; \citet{lis01}, $S>0.4$~Jy at 5~GHz; \citet{all03}, $S>1.3$~Jy at 5~GHz). At these flux densities, the radio source population is dominated by AGN, while at fainter flux densities ($\lesssim$~1~mJy), it is dominated by star-forming galaxies \citep{win99,ric99,hop00}. Research on the variability of faint radio sources has so far yielded small samples as a consequence of the small areas searched and the long integration times required. \citet{car03} observed the Lockman Hole region on timescales of 19 days and 17 months down to 0.1~mJy at 1.4~GHz and discovered nine variable sources, providing an upper limit to the areal density of variable radio sources, as well as constraints on the beaming angle of GRBs and confusion limits relevant to searches for GRB afterglows. Archival VLA calibrator observations spanning 22 years with 944 independent epochs have been used by \citet{bow07} to search for radio transients over a single field with half-power beam widths  of 9\farcm0 and 5\farcm4 at 5~GHz and 8.4~GHz, respectively. The typical observation's flux density threshold is $300$~$\mu$Jy. They detect eight transients in single epochs and two transients in two-month averages of the data. Two of the transients were identified as RSNe, while the absence of optical counterparts for the remainder offers a wide variety of possibilities including Orphaned GRB Afterglows (OGRBA), stellar sources, propagation effects, microlensing events, or perhaps mechanisms heretofore unknown.

In the Galaxy, single radio bursts detected from the Parkes Multibeam Pulsar Survey revealed a new population of neutron stars: Rotating Radio Anomalous Transients \citep[RRATs;][]{mcl06}.  Galactic Center Radio Transients \citep[GCRTs;][]{hym02,hym05,hym09} have diverse light curves with outburst periods and burst durations varying from minutes to months. Lacking counterparts at other wavelengths, their hosts, and the physical mechanisms involved, remain a mystery.

Two large, sensitive, wide-field surveys exist at radio wavelengths. The NRAO VLA Sky Survey \citep[NVSS;][]{con98} covered the entire sky north of -40\arcdeg declination at 1.4~GHz  with a resolution of 45\arcsec and an rms noise of 0.45~mJy. The Faint Images of the Radio Sky at Twenty-cm survey \citep[{\sl FIRST};][]{bec95} has covered $\gtrsim$~9,000~deg$^2$ of the sky at 1.4~GHz with a uniform rms of 0.15~mJy and an angular resolution of 5\farcs4. A number of authors have used these two surveys in their overlap region to search for radio transients. \citet{lev02} tried to constrain the detection rates of orphaned radio afterglows often associated with GRBs. \citet{gal06} attempted to characterize the sample generated by \citet{lev02}. They report the detection of a radio SN in a nearby galaxy and the detection of a source with no optical counterpart, concluding the latter is unlikely to be associated with a GRB. They also place tighter constraints on the beaming factor of GRBs and a limit on total rate of nearby relativistic explosions, implying that most core collapse SNe do not eject unconfined relativistic outflows. These studies have been complicated by the mismatched resolutions and flux density sensitivities of the NVSS and the {\sl FIRST} surveys.

\citet{dev04} have used $\sim$~120~deg$^2$ of the {\sl FIRST} survey data near $\delta=0$\arcdeg taken in 1995 and then repeated in 2002 to study the optical properties of sources that exhibit significant radio variability at 1.4~GHz over this seven-year interval. They find 123 variable objects with flux densities ranging from $\sim$~2~-~1000~mJy. They conclude that there is a higher fraction of quasars in the sample of variables compared to the non-varying sample. 

More recently, \citet{ofe10} have raised a very interesting possibility that the mysterious transients of \citet{bow07} observed out of the plane of the Galaxy, based on their areal number density, duration and energy characteristics, could have a progenitor population consistent with being Galactic isolated old neutron stars with mean distances of the order of kpc. X-ray follow-up \citep{cro11} of the transients reported by \citet{bow07} mostly resulted in non-detections and the X-ray flux upper limits imply consistency of the progenitor population with the above possibility besides being consistent with extreme flare stars at $\gtrsim$~1~kpc or less extreme flares from brown dwarfs at distances of $\sim$~100~pc. 

Here, we use the data from the {\sl FIRST} survey to create the largest, unbiased sample of variable radio sources to date down to a sensitivity level of a few mJy. In \S\ref{FIRST}, we describe the attributes of the {\sl FIRST} survey relevant to a study of radio variability. In \S\ref{dataanalysis}, we present our approach to extracting a sample of variables from the more than two million individual source observations contained in the database. We then describe (\S\ref{selection}) our refinement of the list of variable candidates, the parameterization of their variability, and a summary of their properties. In \S\ref{catalog}, \S\ref{crossid} and \S\ref{summary}, we present the results of our analysis, including the cross-identification of our sample of transients and variables with existing data at optical wavelengths (and in other spectral regimes), and a discussion of the classes of variables identified, as well as the new populations which might exist among the unidentified radio variables.

\section{The Data \label{data}}

The {\sl FIRST} survey, described below, is the most sensitive large-area radio survey in existence, and provides a database that is uniform in angular resolution and flux density sensitivity. Until the next generation of radio instruments arrives, it offers the opportunity to produce the largest unbiased survey yet undertaken for radio transients and variables. We outline here the salient characteristics of the survey. 

\subsection{The {\sl FIRST} Survey \label{FIRST}}

The {\sl FIRST} survey covered $\sim$~9,055~deg$^2$ (8,444~deg$^2$ in the north Galactic cap and 611~deg$^2$ in the south Galactic cap) with observations conducted between 1993 and 2004 using the NRAO VLA in its B-configuration operating at a frequency of 1.4~GHz \citep{bec95}. Roughly 65,000 three-minute snapshot images were obtained over this period. All of the work published to date, including the {\sl FIRST} catalog, has been based on images constructed by co-adding these snapshots with appropriate weightings (\citealp[for details, see][]{whi97}); this process yielded images with a uniform noise level of $\sim$~0.15~mJy. The final catalog contains 816,331 sources with a source detection threshold of $\sim$~1~mJy, yielding a source density of $\sim$~90~deg$^{-2}$. The astrometric accuracy of the cataloged sources is better than 1\arcsec. The co-added maps have 1\farcs8 pixels and an angular resolution of 5\farcs4. The FWHM of the VLA's primary beam is $\sim$~30\arcmin. The catalog also assigns a sidelobe probability to each source. It must be noted that, for this study, we have made use of ``08jul16'' version of the catalog available on the {\sl FIRST} survey's webpage (http://sundog.stsci.edu/first/catalogs/readme.html). 

\citet{bec95} describe the survey design and objectives in full detail.
Here, we provide a brief review of the aspects of the survey relevant to a variability search. In order to obtain uniform sensitivity and sky coverage in the most efficient manner possible, a hexagonal grid is optimal. The pointing centers of each snapshot were placed at the vertices of this hexagonal grid, the shape of which changes slowly with declination. The typical spacing between two neighboring snapshots is $\lesssim$~26\arcmin; thus, all adjacent snapshots overlap. The time between overlapping observations varies: $\sim$~3 minutes between adjacent snapshots along the same declination strip in the east-west direction, $\sim$~1 to many days between adjacent snapshots that lie on two adjacent strips of declination in the north-south direction, and $\sim$~1 to a few years between two adjacent blocks of annual allocations.  Thus, while the survey was not undertaken to find variables and transients, its design is implicitly conducive for a search for radio variability on timescales of minutes to years. 

In this project, we use all of the snapshots from the North Galactic Cap data which covers 8,444~deg$^2$. Part of the South Galactic Cap coverage was repeated, and is in the process of being observed at a third epoch; preliminary variability results have already been reported by \citet{dev04} and a comprehensive analysis of this data set will be undertaken separately when the current observations are completed.

In addition, we have restricted our analysis to isolated point sources. From light travel-time arguments, extragalactic sources that vary significantly on timescales of minutes to years are necessarily compact. Even though compact components of extended sources could vary, we limit our attention herein to point-like (unresolved) sources with a sidelobe probability below 0.15. Further, an additional selection criterion was imposed requiring these sources to be relatively isolated, with no neighbors within 18\arcsec; this minimizes the likelihood that differing multi-component fits between observations will produce spurious evidence of variability.

\begin{figure} \centering
\includegraphics[width=\linewidth]{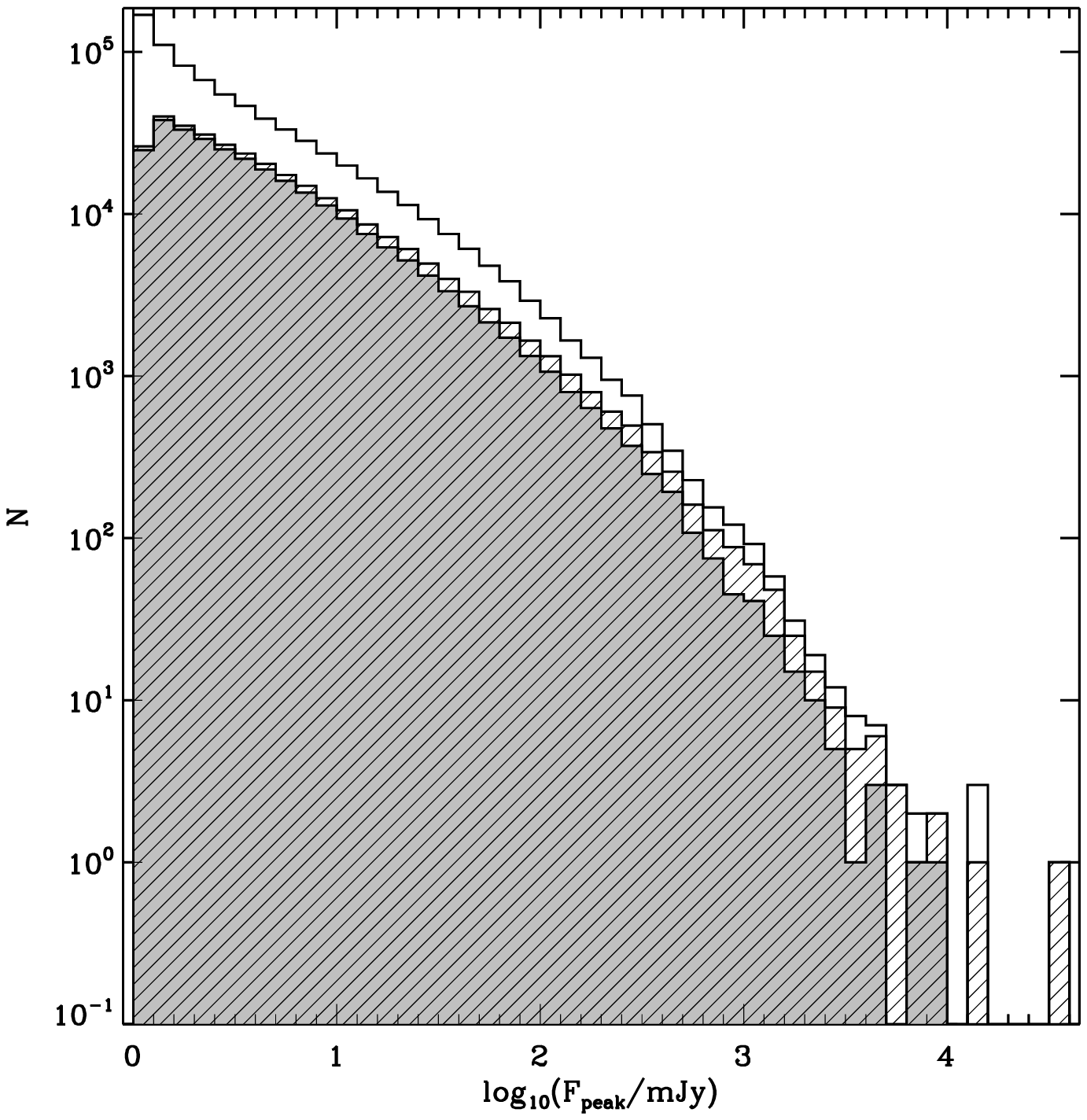}
\caption[Histogram of peak flux densities for sources in the {\sl FIRST} survey]{Histogram of peak flux densities for sources in the {\sl FIRST} survey. 
The hatched area includes all point sources, while the black shaded
histogram is for the isolated point sources used in this study.\label{FIRSTflux}}
\end{figure}

Figure~\ref{FIRSTflux} shows the flux density histograms of all sources, point-like sources, and isolated point-like sources. 
Out of a total source count of 758,942 in the Northern Galactic Cap, there are 647,550 sources with sidelobe probabilities less than or equal to 0.15. Of these, 519,537 are isolated sources and 279,407 meet our criterion for being point-like (major axis $\leq$~5\farcs97 for $\delta\leq$~4\arcdeg~ declination and major axis $\leq$~6\farcs87 for $\delta >$~4\arcdeg~ declination). We examine the variability characteristics of these 279,407 sources which have been observed between two and seven times each.

\subsection{VLA Data-recording Glitch \label{VLAglitch}}

It recently became clear that the on-line VLA recording system suffers from a rare glitch that manifested itself many times in the {\sl FIRST} snapshot database. Although the exact nature and cause for this glitch have not been completely understood, the result is that a few visibilities from one snapshot can be appended to the adjacent snapshot. This issue has also been pointed out by \citet{ofe10}. The prototypical example that led to the discovery of this glitch is the source claimed to be a transient based on a comparison with the NVSS: VLA 172059.9+38522 \citep{gal06,lev02}. The authors speculated that this could be a prototype for a new class of transients. Upon investigation, it was discovered that one of the snapshot images containing the location of this source had been affected by the VLA glitch. A thorough search of the database found that 190 snapshot images (0.29\%) are apparently affected by this problem; while in many cases the problem can be fixed by deleting a few records from the $uv$~data set, in other cases subtler anomalies appear. As a consequence, we have simply excluded all these snapshot images from our analysis. Additional grid images excised include those found to contain imaging artifacts and those with large calibration errors (see below). 

\section{Data Analysis \label{dataanalysis}}

Using the positions of isolated, point-like sources from the {\sl FIRST} catalog, we created a comprehensive inventory of each object location and the corresponding pixel locations in each of the grid images in which it appeared. Every object was found to be covered by from two to seven snapshot images as expected from the survey design described above. 

\subsection{Gaussian Fitting \label{gaussfit}}

In each of the snapshot images, at the location of the object, a peak flux density was estimated by fitting a Gaussian plus a flat baseline component\footnote{For the purposes of this paper, it should be noted that our use of the term ``baseline'' refers to a flat or constant component in the Gaussian fitting and should not be confused with the baseline vectors (in units of wavelengths) often used in radio interferometry}. The position, width and position angle of the Gaussian, already available from the catalog, were held fixed. The RMS noise in each image was also measured around the object's location. For every pixel in each image, the "local" standard deviation was calculated by considering a box four times the FWHM of the synthesized beam centered on that pixel (the FWHM is three pixels). Assuming an object of interest is located at a given pixel, we consider a circle consisting of pixels at radius 5 times the FWHM of the synthesized beam. Thus, each of the pixels in this circle has a standard deviation that was effectively calculated using pixels in an annulus that has roughly an inner radius of 3 times and an outer radius of 7 times the FWHM of the synthesized beam. We associate the median of all these standard deviations with the rms noise around the object. The advantage of using a median is it effectively ignores pixel intensities that are not purely noise such as those from another source (or a sidelobe) near to the object of interest. In the case of a marginal (or non-) detection of the object in a snapshot image, a $3\sigma$ upper limit was assigned. 

\subsection{Identification \& Removal of Striped Images \label{HiBase}}

Artefacts in radio interferometric images arise for a number of reasons such as calibration errors, interference, etc. For a large-scale survey such as {\sl FIRST}, it is difficult to ensure high data quality at all times during the observations. One not uncommon problem is a striped pattern corrupting an image owing to un-excised interference or other deconvolution problems.

\begin{figure} \centering
\subfigure[Before flagging]{
  \includegraphics[width=0.475\linewidth]{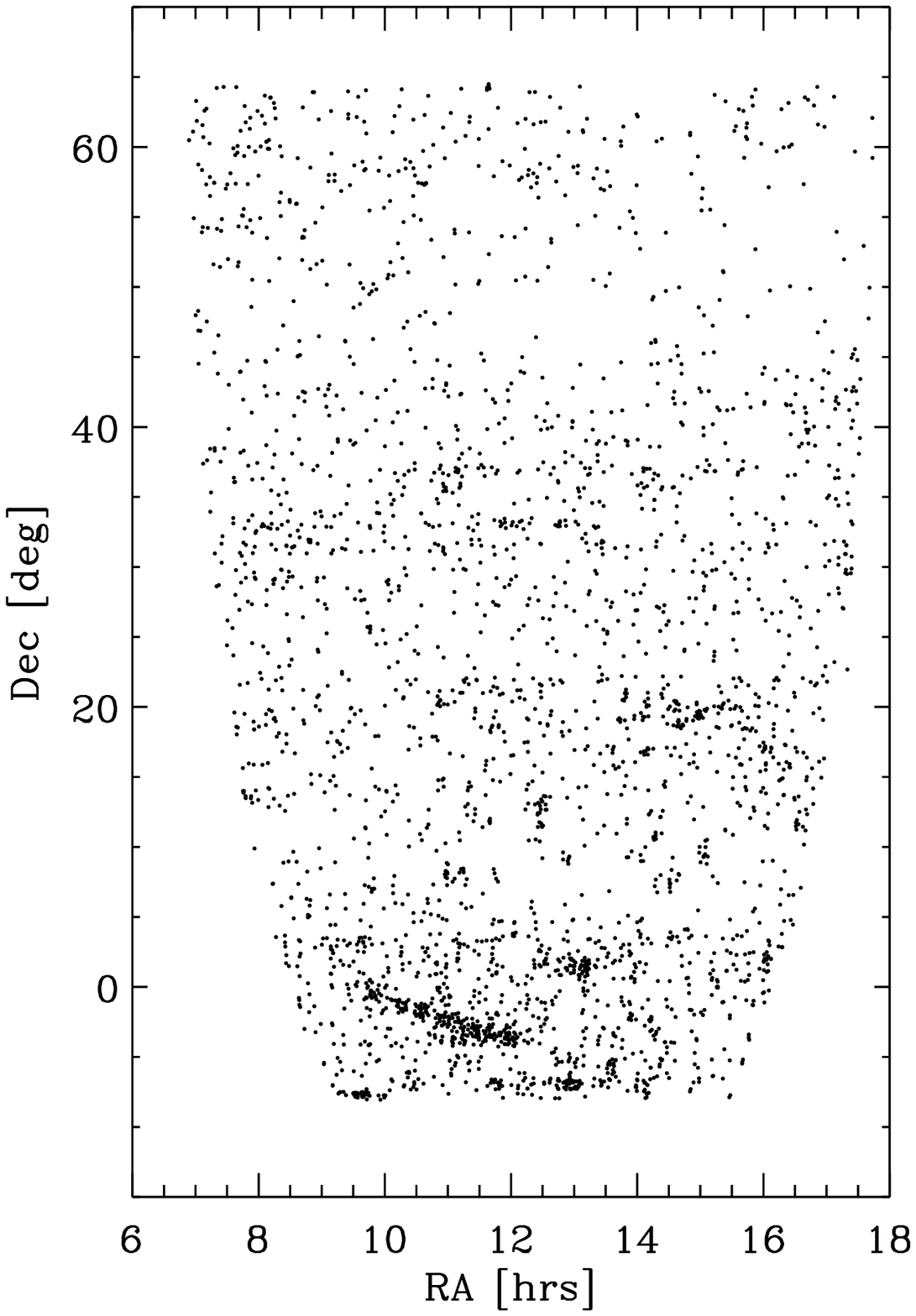}}
\quad
\subfigure[After flagging]{
  \includegraphics[width=0.475\linewidth]{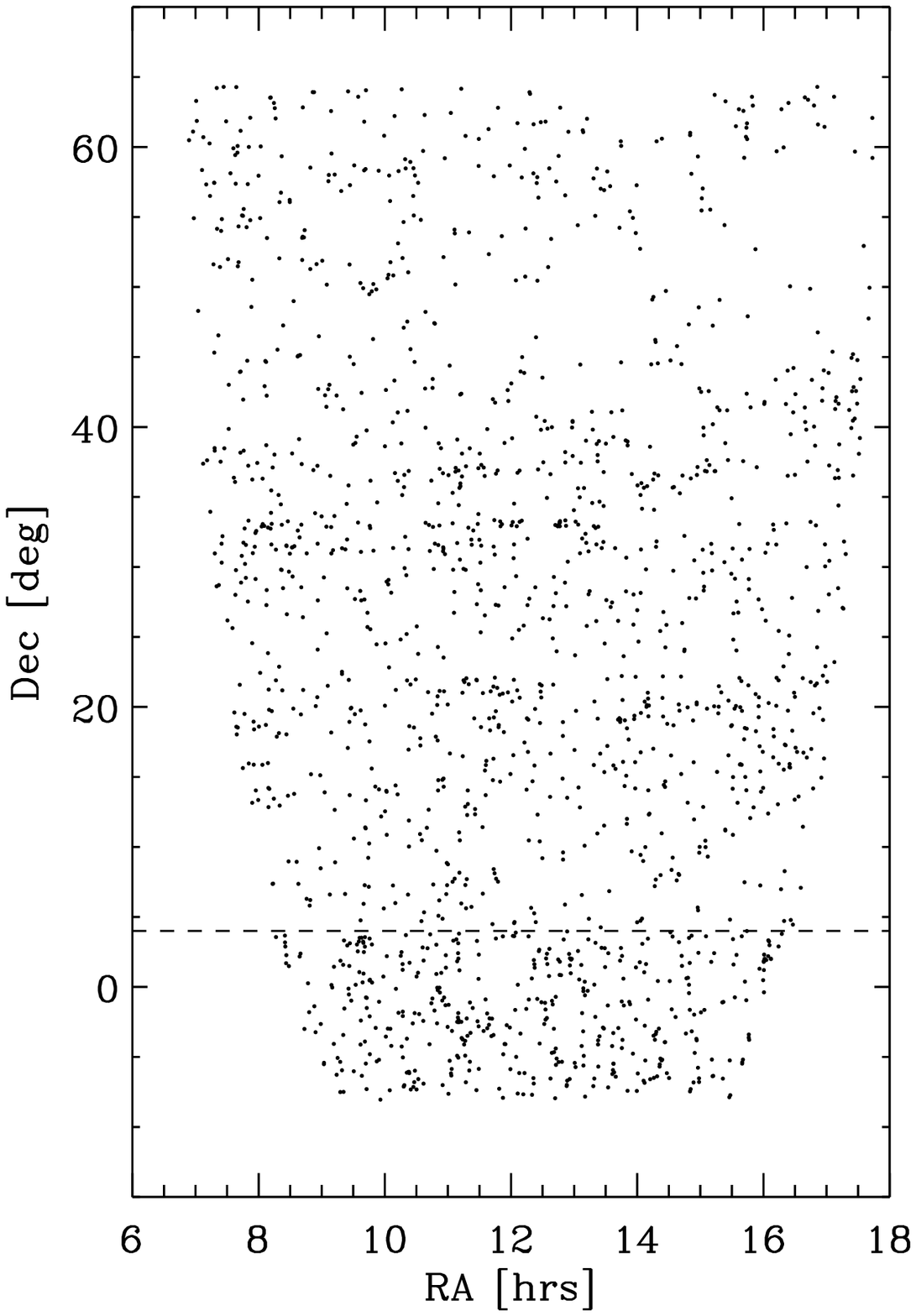}}
\caption[The sky positions of outliers before and after flagging and removing the snapshot images that contain one or more sources with a significant baseline component from the gaussian fitting routine]{The sky positions (J2000) of outliers before ({\it Left}) and after ({\it Right}) flagging and removing the snapshot images that contain one or more sources with a significant baseline component from the gaussian fitting routine (See \S\ref{HiBase}). Note the clear reduction in the source clusters.  The horizontal dashed line indicates the boundary in declination below which the outlier density shows enhancement; this is a consequence of the change in the convolving beam size which reduces the number of sources eliminated from consideration by modest extent (cf. Figure~\ref{dec_PS_size}).\label{sky_positions}} 
\end{figure}

After deriving a preliminary list of candidate variables, we plotted their positions on the sky, and found that some objects were spatially clustered. These clusters had different shapes, including linear, elliptical and irregular appearances, and their sizes spanned several degrees on the sky. Such strong clustering is highly suggestive of systematic problems with the images. We investigated the snapshot images associated with these clustered objects and found that many, especially toward the southern limit of the survey, have clearly visible striped patterns running across them. The width of these stripes is typically greater than the angular resolution of the images. Having restricted ourselves to isolated point-like sources, we were able to make use of the baseline fitting described in \S\ref{gaussfit}. We flagged any image that contained one or more sources with a baseline component greater than 1.75 times the local image noise. A total of 2162 (3.96\%) such snapshot images were deleted from subsequent analysis. Figure~\ref{sky_positions} illustrates how this step significantly eliminated the non-physical clustering of variable and transient source candidates.

\subsection{Identifying Global Amplitude Calibration Errors \label{MWRAT} }

While amplitude and phase calibration errors are minimized in the {\sl FIRST} imaging pipeline for individual snapshots, it is important to keep in mind that the survey is spread over very large scales in space and time. The overall calibration scale, then, could well vary over the course of the survey \citep[see][]{whi97}. Relevant to our study is the variation of the overall amplitude calibration scale over the tens of thousands of snapshot images.

An estimation of the calibration error was performed after a first iteration of steps in \S\ref{empiricalpbeam} but before the identification of any variables or transients. The algorithm used considers sources in each snapshot image and compares their behaviour to that in all the neighbouring snapshots in which they are also found. The null hypothesis, true for a vast majority of the sources, is that sources will exhibit a constant peak flux density across the different snapshots in which they are detected. A best-fit algorithm assigns a multiplicative amplitude calibration correction factor to each snapshot image such that, when it is applied, all the sources in the neighborhood are as close to the null hypothesis behavior as possible in a minimum absolute deviation sense. Thus, the amplitude calibration variations are estimated locally. 

\begin{figure} \centering
\includegraphics[width=\linewidth]{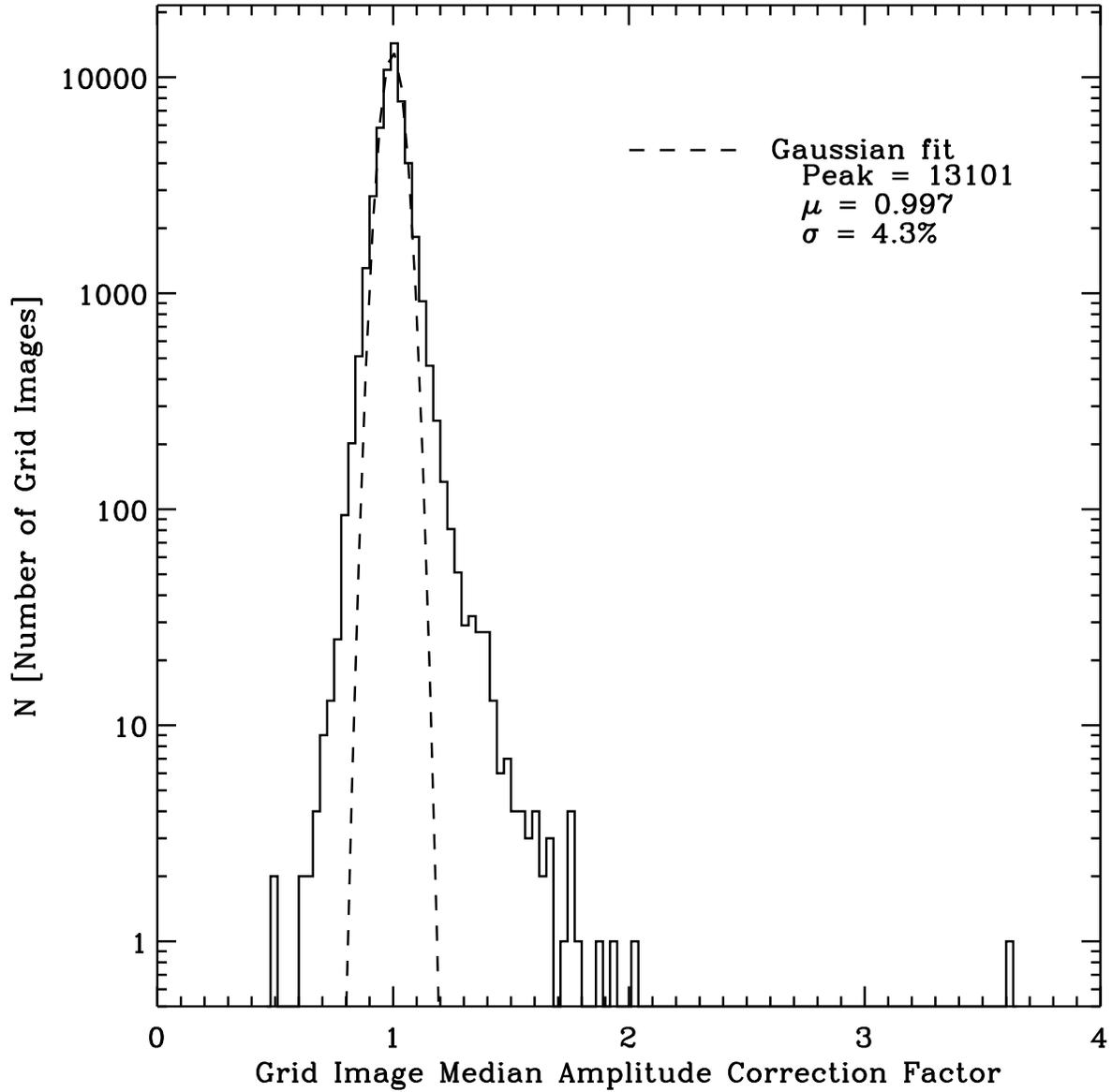}
\caption[Histogram of overall amplitude correction factors for grid images]{Histogram of overall amplitude correction factors for grid images. The dashed line represents the best gaussian fit to the histogram whose parameters (peak, mean ($\mu$) and standard deviation ($\sigma$)) are displayed in the plot.\label{MWRAThist}}
\end{figure}

Figure~\ref{MWRAThist} shows the histogram of the amplitude correction factors, where a value of unity indicates no correction need be applied. It can be seen that the distribution, especially the inner core comprising the large majority of snapshot images, can be reasonably fit with a gaussian profile with a standard deviation of 4.3\%. However, tails can be seen on both sides of the distribution indicating some large fluctuations in the amplitude calibration for a small fraction of the images. We have accounted for the typical calibration uncertainty using a conservative value of 5\% in determining the flux density uncertainty for individual snapshots as described in \S\ref{errors_contrib}. 

\begin{figure} \centering
\includegraphics[width=\linewidth]{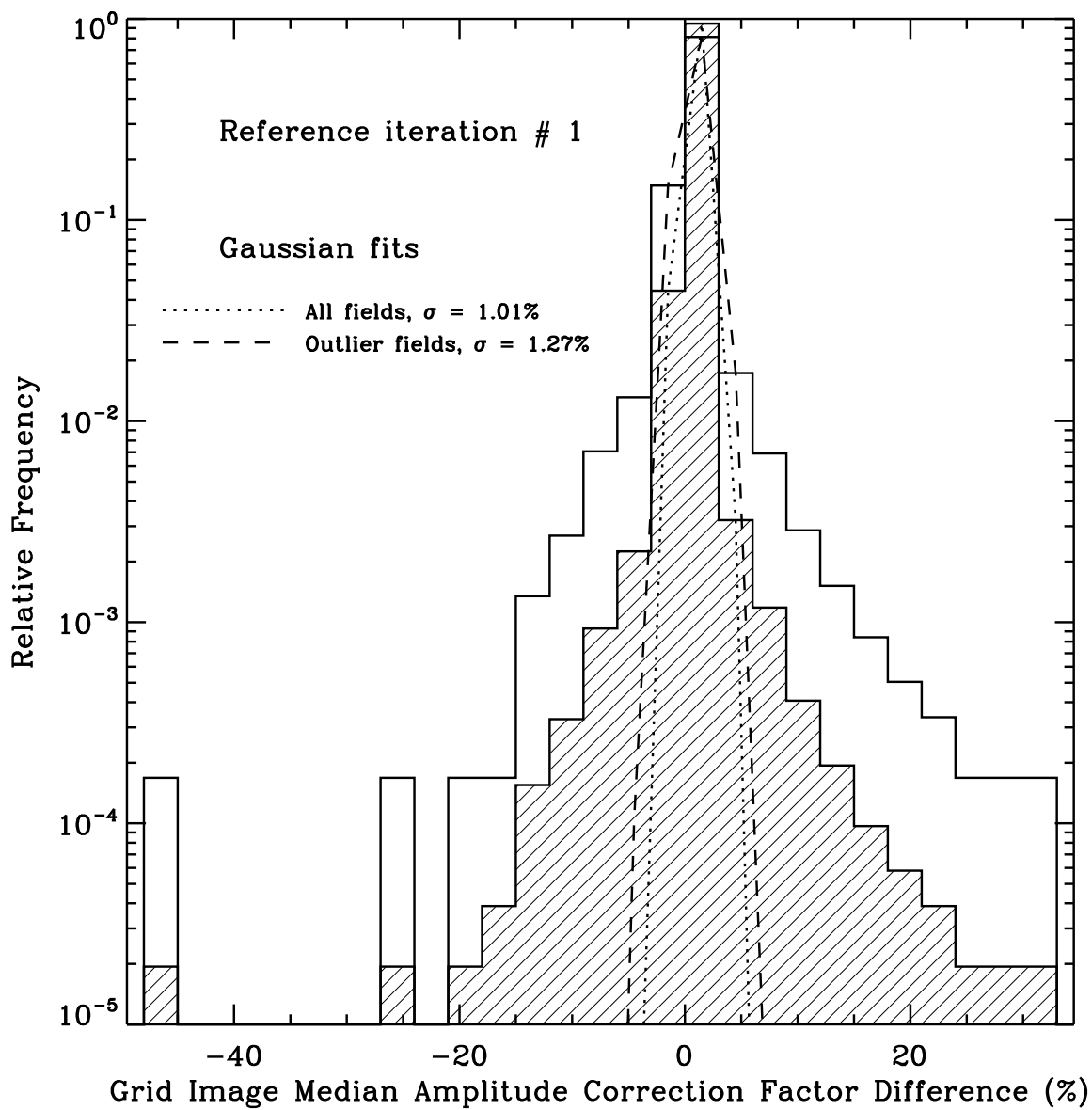}
\caption[Relative frequencies of the difference in median calibration offsets for the grid images between the first and second iterations]{Relative frequencies of the difference in median calibration offsets (in percent) for the grid images between the first and second iterations where the second iteration has excluded the outliers (determined from the previous iteration) from the calibration calculations. The shaded portion represents all snapshot images while the unshaded portion represents snapshot images that contain one or more outliers. The unshaded distribution is clearly wider than the shaded one as indicated by the gaussian fits, implying that the amplitude calibrations of the snapshot images are affected by the inclusion of the outliers.\label{MWRATiter}}
\end{figure}

We experimented with different thresholds for the largest allowable amplitude calibration offsets. We set the threshold at 20\%; for amplitude offsets below this value, the number of candidate variables eliminated was proportional to the number of images eliminated, whereas snapshots with higher amplitude corrections contained disproportionally larger numbers of candidates. We also performed an extra iteration, computing the amplitude correction factor after removing sources that were found to be candidate variables and recomputing the calibration correction factors. We find evidence that the change in the amplitude correction factors between the two iterations is greater for snapshots that contain putative variables than for ones that do not (figure~\ref{MWRATiter}). However, the final list of variable candidates between one such iteration and the next changes by only $\sim$~2\%. Thus, we have used only snapshot images that have an amplitude offset less than 20\% after one iteration of the amplitude correction factor estimation.

For the set of images retained, we apply, on an image by image basis, the overall amplitude calibration correction factors to the peak flux densities, local image noise values, and baseline components for all the sources found in those images. Hereafter, these corrected values are treated as the true values.

\subsection{Determining the Primary Beam Empirically \label{empiricalpbeam}}

The data were obtained in 2$\times$7 3-MHz channels near 1.4~GHz. While the center frequency of these bands was fixed for most of the survey observations, the interference environment was constantly changing, leading to data editing which affects the effective center frequency of the observing band(s). Thus, the frequency-dependent primary beam, normally derived from the AIPS task PBCOR, cannot be adopted blindly. 

Our selection algorithm makes use of the hypothesis that the non-varying source population is expected to dominate the variable one. Hence, with the off-axis angle information available for each source in each snapshot image, the primary beam can be determined empirically. We see from figure~\ref{PBEAM} that the difference between our estimate for the primary beam and the nominal VLA primary beam is at most a few percent. As figure~\ref{errors} demonstrates, the uncertainty arising from the primary beam estimation is a negligible contribution to the overall error budget, independent of source flux density.

The peak flux density of each isolated, point-like object measured in each snapshot image (as described in \S\ref{gaussfit}) is normalized by the reported catalog flux density \citep[after taking into account the CLEAN bias as quantified by][]{whi07}. Since the snapshot images and the fitted peak flux densities derived therefrom have not been corrected for the primary beam attenuation, adopting the null hypothesis that a source's flux density stays constant means that these normalized values yield a primary beam normalization at the off-axis angle of the measured source. When this normalization is performed for all objects, we obtain an empirical primary beam:

 \begin{displaymath}
   \mathrm{empirical\; PB} = \frac{\mathrm{measured\; map\; peak\; flux\; density\; (mJy)}}{\mathrm{catalog\; peak\; flux\; density\; (mJy)\; -\; CB\; (mJy)}}
 \end{displaymath}

 where CB is the CLEAN bias \citep{whi07}.

The empirical primary beam values obtained from all the objects in each of their corresponding snapshot images were binned in angular offset from the center of the primary beam. A sixth-degree polynomial was fitted to determine an analytical expression for the empirical primary beam. The error in the primary beam estimate is the uncertainty in the mean estimate for the primary beam in each bin, which is the standard deviation in each bin divided by the square root of the number of sources in that bin. Thus, a primary beam error estimate (used in \S\ref{errors_contrib}) is assigned to each data point on each light curve depending on which bin of angular offset the data point falls into. 

Figure~\ref{PBEAM} shows the scatter plot of normalized flux densities from all the usable snapshot images in the upper panel; in the lower panel, the bin-averaged data points from the upper panel are overplotted on the polynomial fit. Fits were obtained independently in eight separate declination zones, but no significant variations ($>$~0.9\%) with declination were seen, so a single empirical primary beam was used for the whole survey.

\subsection{Sources of Uncertainty in the Peak Flux Density \label{errors_contrib}}

We next estimate the sources of uncertainty in the peak flux density obtained for an object detected in a snapshot image. If $F = Af/p$, where $F$ is the reported flux density, $A$ is the absolute amplitude calibration (ideally $A=1$) and $p=p(\theta)$, the primary beam value at an offset ($\theta$) from the field center, the principal uncertainties that need to be accounted for include the following:

  \begin{enumerate}

    \item local noise in the grid image in the vicinity of the source ($\Delta f$):  This will further be amplified because of the primary beam effect, especially for sources far from the field center. $\Delta F = A\,\Delta f / p$.

    \item uncertainties in primary beam ($\Delta p$): We have computed empirically the primary beam and its uncertainty from the data as described in \S\ref{empiricalpbeam}. $\Delta F = A\,f/p^2\,\Delta p$.

    \item uncertainties in the overall flux calibration amplitude ($\Delta A$): This uncertainty is $\sim$~5\% and was estimated as described in \S\ref{MWRAT}. $\Delta F = \Delta A\,f/p$.

    \item pointing errors of the VLA antenna ($\Delta \theta$): This is not the same as the positional accuracy of the sources in the catalog. The antennas have an rms pointing error of $\Delta \theta \sim$~10\arcsec. This translates to an error in flux density, $\Delta F = A\,f\,(\mathrm{d}p/\mathrm{d}\theta)\,\Delta \theta/p^2$. 

  \end{enumerate}  

  The overall error is obtained by adding all the above errors in quadrature. The dominant contributions come from the map noise for fainter sources and from the amplitude calibration errors for brighter sources. 

  \begin{displaymath}
    \Delta F = \sqrt{\left(\Delta A\,f/p\right)^2 
      +\left(A\,\Delta f/p\right)^2
      +\left(A\,f/p^2\,\Delta p\right)^2
      +\left(A\,f/p^2\,\frac{\mathrm{d}p}{\mathrm{d}\theta}\,\Delta \theta\right)^2}
  \end{displaymath}

Figure~\ref{errors} shows the contributions of these different sources of uncertainty as a function of angular offset from the center of the image for two sources observed (without primary beam correction) to have flux densities of 1~mJy and 10~mJy, respectively.

Three other potential sources of error have been assessed and found to be negligible compared to the above terms. Firstly, we adopt for each source the FIRST catalog position and hold it fixed when fitting for a source's peak flux density in each grid image. Any systematic astrometric errors which are a function of position in the primary beam field of view will thus induce an underestimate of source flux density. We have assessed such errors and report the results for the co-added images in \citet{mcm02}. We have also assessed the errors for the grid images and find similar results: the maximum error is $<$0\farcs3. This will induce a a maximum flux density error of $<1\%$.

A second potential effect involves the (generally unknown) source spectral index. Since the primary beam shape is a function of observing frequency, the primary beam correction changes from the lower to the upper end of the observing band. While our empirical determination of the primary beam correction assures that the average value is correct for the average source spectral index, a source with an extremely steep or inverted spectrum will have an inappropriate primary beam correction applied. We have evaluated the magnitude of this effect for extreme sources with spectral indices of $+2.0$ and $-1.5$ compared to a mean source spectral index of $\sim -0.7$. At an off-axis angle of 20\arcmin the error is less than 1\% in all cases. At 30\arcmin off axis, it is $\lesssim 2\%$ for $-1.5<\alpha<+1.0$; for an extremely inverted (and very rare) source with $\alpha=+2.0$ the error does reach 4\%, but the other errors rise rapidly this far off-axis and still dominate the total error budget.

Finally, we address the matter as to whether or not variations in the small CLEAN bias added to each measured flux density could generate spurious variability. CLEAN bias is an at-best partially understood manifestation of the non-linear CLEANing algorithm applied to snapshot data \citep[see][]{bec95,con98}. It is known to vary with source extent and map rms; in principle, it could also vary with time. For the FIRST catalog, artificial source experiments (inserting sources of known flux density into a $uv$~data set and then measuring the flux density recovered in the derived image) suggested that, for the point-like sources considered here, a constant value of 0.25~mJy provided a satisfactory correction. Becker et al. (1995) did note that for fields with sources brighter than 500~mJy (and thus higher rms values), the bias correction could be as great as 0.50~mJy leading to a 0.25~mJy underestimate of the true flux density. All variable candidates in such fields discovered here are flagged. 

Subsequently, we examined this issue in \citet{whi07} using artificial sources as well as an analysis of sources from the deep radio imaging of the Spitzer First-Look Survey \citep{con03}. For 145 sources from different fields in the flux-density range 1 to 3~mJy, the CLEAN bias in FIRST images was shown to be $0.25 \pm 0.04$~mJy. Thus, even for 1~mJy on-axis sources, the size of any CLEAN-bias error is less than that of image noise and the amplitude correction error; for brighter, and/or off-axis sources, the uncertainty is negligible.

\section{Selection of Variable Candidates \label{selection}}

\subsection{Mean Inferred Peak vs. Catalog Peak Flux Density \label{Fmean_Fpeak}}

Owing to the reasons discussed in \S\ref{VLAglitch}, \S\ref{HiBase} and \S\ref{MWRAT}, the snapshot images that form the basis of our search are a subset of those from which the {\sl FIRST} catalog was constructed. The catalog peak flux densities were derived by a weighted average estimate of the inferred fluxes where the weights are related to the square of the primary beam pattern \citep{whi97,con98}. We have recalculated the mean value for each source using only those snapshot images actually included in our search and denote it by $\overline{f}$, where $\overline{f} = \sum_i{p_i^2f_i}/\sum_i{p_i^2}$. $\overline{f}$ and $\langle f\rangle$ will be used interchangeably hereafter. 

Inferred peak flux density points that lie far from $\overline{f}$ are considered outliers. We have selected as candidate variables outliers that deviate from $\overline{f}$ by more than five times their inferred peak flux density uncertainty, defined in \S\ref{errors_contrib}. 

\subsection{Removal of Contamination from False Positives \label{false_positives} }

We have identified several additional effects other than genuine variability that can generate outliers.

\begin{figure} \centering
\includegraphics[width=\linewidth]{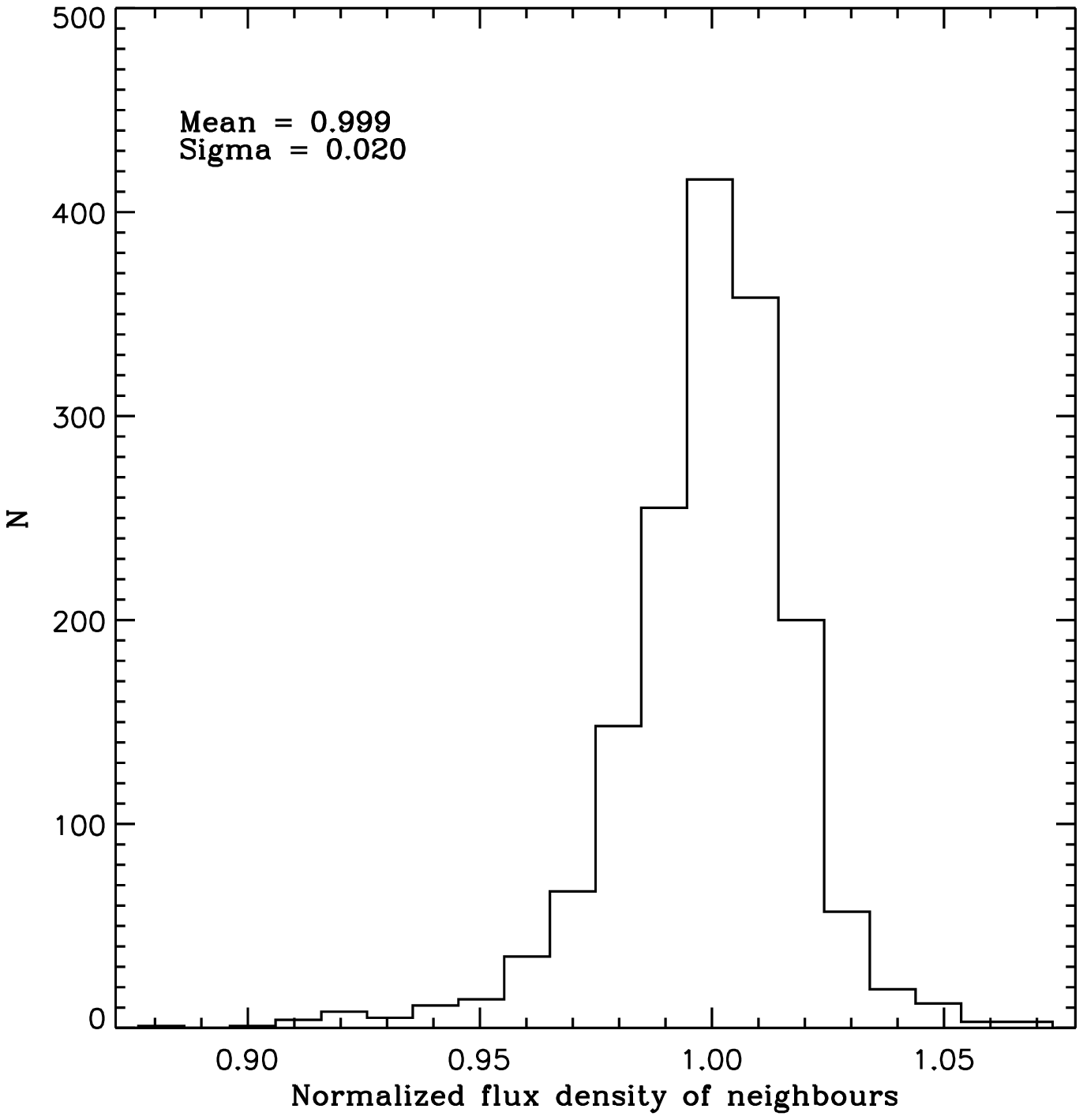}
\caption[Histogram of mean normalized flux densities of neighbouring sources associated with outliers]{Histogram of mean normalized flux densities of neighbouring sources associated with outliers.\label{nghbrs_behaviour}}
\end{figure}

\begin{itemize}

\item \textbf{Low-Quality Images}: Noisy images and images with interference and other deconvolution errors are significant sources of false detections. Such outliers have been largely eliminated using the technique described in \S\ref{HiBase}; a few additional bad fields containing large numbers of spurious sources were also deleted.

\item \textbf{Sidelobes}: There were instances in which sidelobes from
  nearby sources have significant flux density in one grid image but are
  absent or marginal in others, thus producing an apparent source that  
  looks both genuine and variable. The {\sl FIRST} catalog-generation algorithm does calculate a sidelobe probability for each source, and we selected only objects with a sidelobe probability less than 0.15. Nonetheless, some unlabeled sidelobes are still present. All fields containing bright sources were examined carefully to eliminate sidelobes as spurious variables. Variable and transient sources suspected of being sidelobes from strong sources with peak flux densities of $>$~500~mJy within a radius of 31\arcmin \, have been flagged. 

\item \textbf{Calibration Problems}: Defective calibration can cause sources to vary between different snapshot images even after incorporating the grid image amplitude correction factors. To identify and eliminate such sources, we used the catalog to select a few (usually between two and four) objects that are neighbors of each outlier (within a 9\arcmin \, radius) and are also point-like. These were normalized by their respective mean peak flux densities. Each was examined for any deviations from the ideal normalized value of unity. Some neighbors exhibited the same  variability pattern as the outliers, thus confirming the presence of remaining calibration problems. Some others were simply inconsistent with no
variation, making the corresponding outliers suspect.  We used this procedure on all outliers to select a final list that appeared to be free from calibration problems. About 30 outliers have neighbors with a deviation from the mean of $3\sigma$ ($\sim 5$ would be expected from a  normal distribution), and 17 have a value $>3.5\sigma$ (figure~\ref{nghbrs_behaviour}); the table includes the deviation for the neighboring sources in sigma for each source. 

\end{itemize}

\begin{figure} \centering
\includegraphics[width=\linewidth]{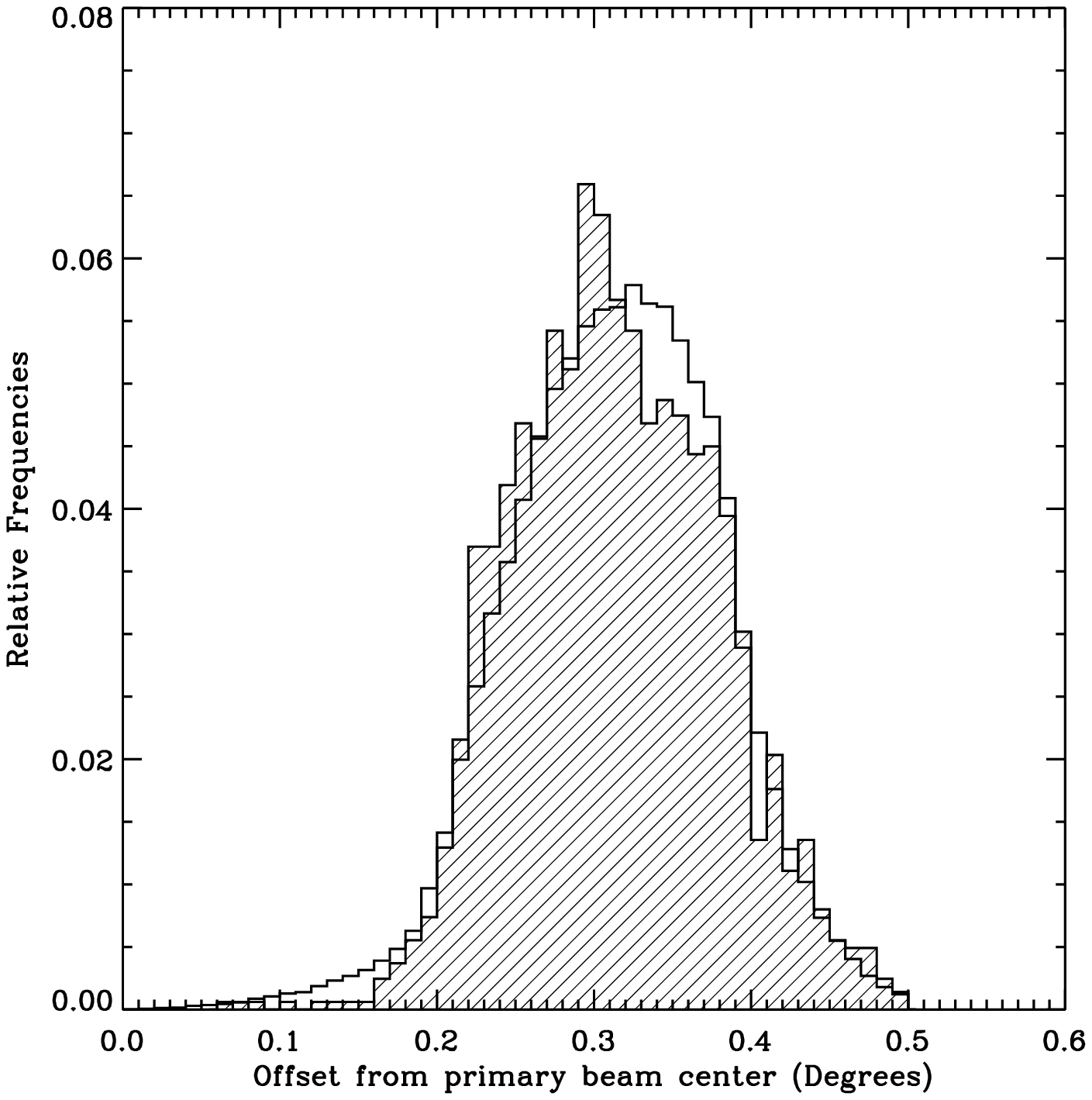}
\caption[Relative frequencies of the off-axis angles of the most discrepant data point from $\langle f\rangle$]{Relative frequencies of the off-axis angles of the most discrepant data point from $\langle f\rangle$. The unshaded distribution corresponds to all the isolated point-like sources in {\sl FIRST} while the shaded distribution corresponds to the variables and transients in the sample.\label{hist_offset}}
\end{figure}

We also checked to see if there is any systematic variation of the fraction of variables and transients with off-axis angle. We plotted the relative frequencies of the off-axis angle for the source in the light curve containing the data point most discrepant from the mean peak flux density. Simulataneously, we plotted the same quantities for all the isolated point-like sources in {\sl FIRST}. In figure~\ref{hist_offset}, we find no major systematic differences between the outliers and the rest of the isolated point-like sources in {\sl FIRST}. The slight shift of the distribution of variables toward the field center arises from the fact that the low {\it observed} flux densities for sources far off axis increases their uncertainty, and thus reduces our ability to detect variability.

After removal of all identified false positives, we are left with a sample of 1627 variables and transients.

\subsection{Indicators of Variability \label{variability_measures}}

We used three indicators to select variable and transient candidates.

\subsubsection{$\chi^2$-probability \label{chisq_prob}}

 We define $\chi^2 = \sum_i \left(\frac{f_i-\overline{f}}{\sigma_i}\right)^2$ and $\chi^2$-probability $=P(\chi^2,\nu)$ as the probability that a $\chi^2$-statistic falls above a certain value $\chi^2 = \sum_i \left(\frac{f_i-\overline{f}}{\sigma_i}\right)^2$, with $f_i = $~inferred peak flux density of the $i^\textrm{th}$ data point, $\overline{f} = $~average inferred peak flux density weighted by the primary beam, $\sigma_i = $~uncertainty in $f_i$, $\nu=n-1$ is the number of degrees of freedom, and $n$, is the number of data points in the light curve. Since there is only one constraint in the model for the null hypothesis -- that the peak flux density is constant with time -- it follows that $\nu = n-1$.

 The probability threshold chosen is $P(\chi^2,\nu)\leq 5.733\times 10^{-7}$ which is equivalent to the probability that the absolute value of a normally distributed variable falls beyond five standard deviations. 

\subsubsection{Maximum Deviation from the Mean Inferred Peak Flux Density \label{best_sigma_from_Fpeak}}

An alternate indicator of variability is the maximum deviation of any individual data point in a light curve from the null hypothesis that the flux density is constant at a value given by the mean inferred peak flux density. We denote this by $\sigma_\textrm{max}$ defined as $\textrm{MAX}\left\{\left|\frac{f_i-\overline{f}}{\sigma_i}\right|\right\}$. We chose a threshold $\sigma_\textrm{max}\ge 5$. 

\subsubsection{Data Point Pair with Most Significant Flux Density Difference \& Timescale Information \label{maxSNR}}

In order to preserve information about variations between any pair of data points in a light curve (and the associated timescale), we also estimate the variation of each point from all the others in a light curve. The significance of this difference is given by the ratio obtained by dividing the difference by the uncertainties in the two flux density estimates added in quadrature, 
   \begin{displaymath}
     \Delta_{ij} = \frac{\left|f_i-f_j\right|}{\sqrt{\sigma_i^2+\sigma_j^2}}
   \end{displaymath}
   where, $\Delta_{ij} = $~significance ratio between $i^\textrm{th}$ and $j^\textrm{th}$ data points, $f_i = $~inferred peak flux density at the $i^\textrm{th}$ data point, $f_j = $~inferred peak flux density at the $j^\textrm{th}$ data point, $\sigma_i = $~uncertainty in $f_i$, and, $\sigma_j = $~uncertainty in $f_j$. We select the pair that has the maximum absolute value defined as $\Delta_\textrm{max}=\textrm{MAX}\left\{\Delta_{ij}\right\}$. For an outlier selected with this indicator, the threshold is $\Delta_\textrm{max}\ge 6$.

$P(\chi^2,\nu)$ is an indicator of overall variability of a light curve that can arise from low but sustained variations, large, sudden variations, or both. Although $P(\chi^2,\nu)$ provides a valid statistical indicator of variability, information about timescales and the temporal localization of the variability has been ignored. Unlike the indicator $P(\chi^2,\nu)$, $\sigma_\textrm{max}$ and $\Delta_\textrm{max}$ are indicators of large sudden variability and temporally localize the variability in the light curve.

A similar estimation of $P(\chi^2,\nu)$, $\sigma_\textrm{max}$ and $\Delta_\textrm{max}$ are performed for the neighbours of the outliers. The respective thresholds were chosen as those beyond which the population of varying sources and their non-varying neighbours look statistically different as defined by a K-S test.

We found that the variability measure given by $\Delta_\textrm{max}\geq 6$ identified no new variables that were not already found by one of the other two measures, namely, $P(\chi^2,\nu)\leq 5.733\times 10^{-7}$ and $\sigma_\textrm{max}\ge 5$. However, we include this variability measure in subsequent discussions because it represents a useful quantity in describing the most significant flux density difference between data points in a light curve and provides us with an idea of the timescale of variability.

\begin{figure} \centering
\includegraphics[width=\linewidth]{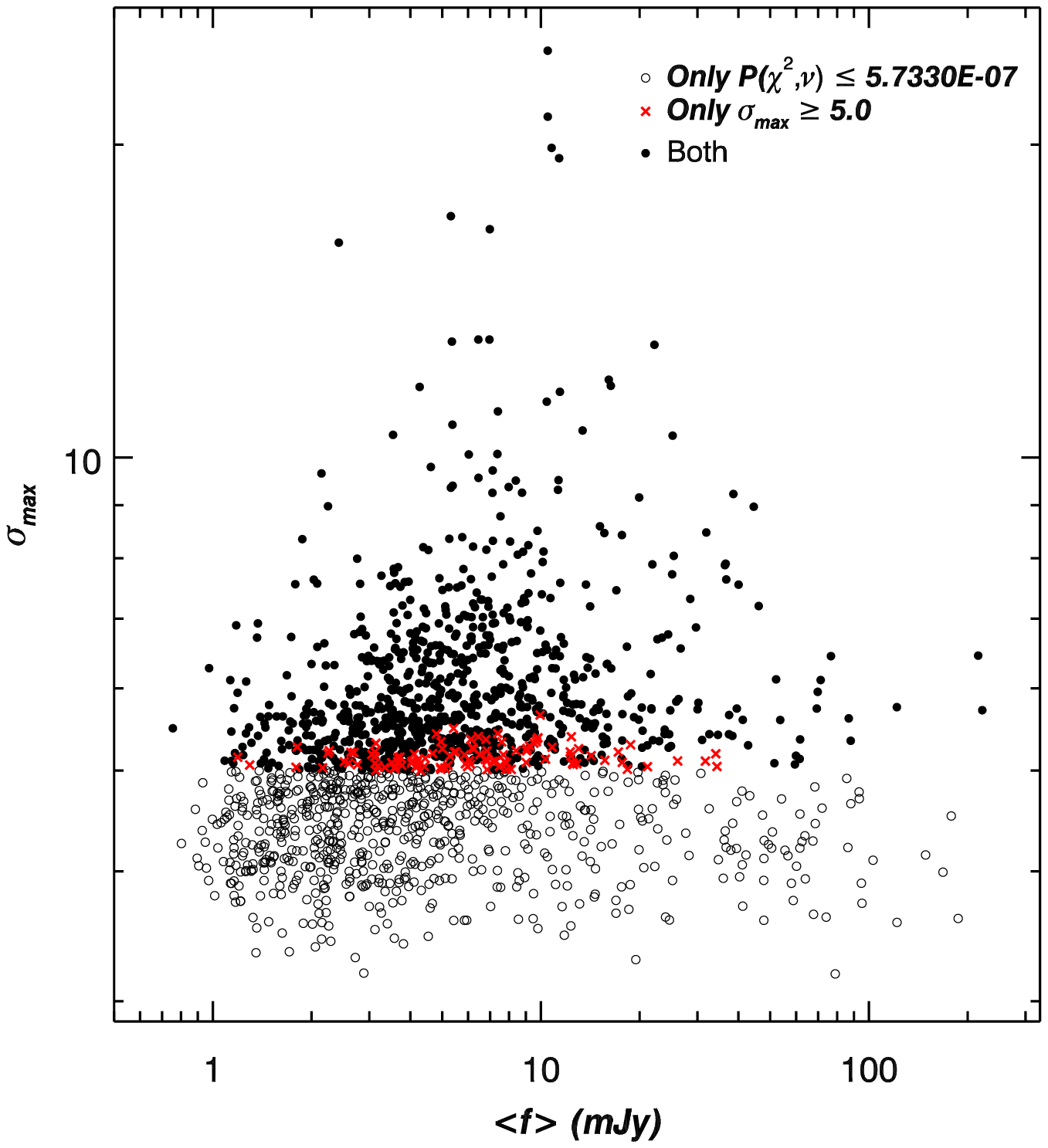}
\caption[The measure of variability, namely, $\sigma_\textrm{max}$ vs. average peak flux density]{The measure of variability, namely, $\sigma_\textrm{max}$ plotted against average peak flux density. The different symbols correspond to outliers that were selected on the basis of various measures of variability.\label{scatter_plot_best_sigma_mean_Fpeak}}
\end{figure}

\begin{figure} \centering
\includegraphics[width=\linewidth]{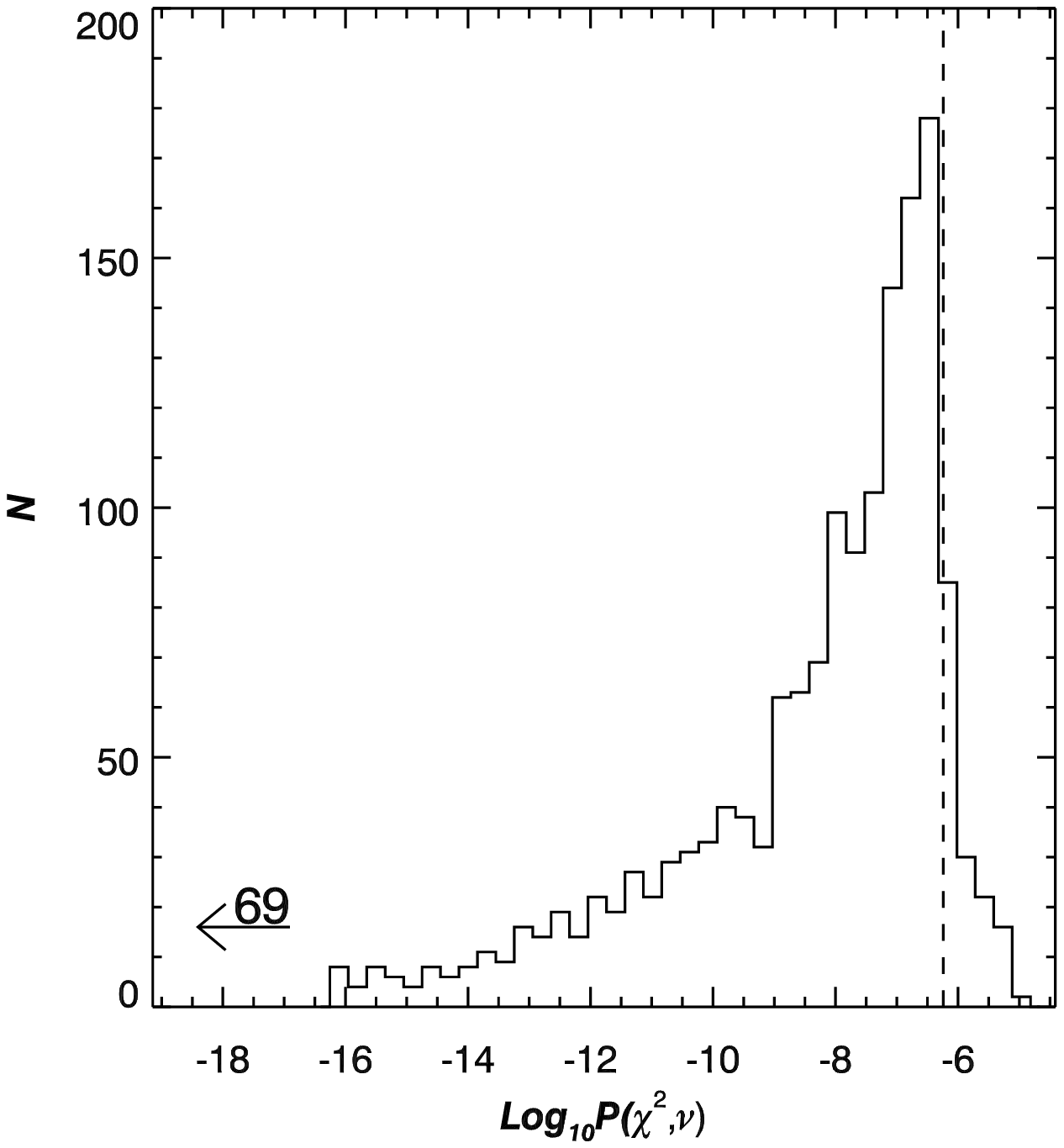}
\caption[The histogram of logarithmic values of $\chi^2$-probability, $P(\chi^2,\nu)$]{The histogram of logarithmic values of $\chi^2$-probability, $P(\chi^2,\nu)$. The fraction of outliers to the right of the dashed line are those selected by criteria other than $P(\chi^2,\nu)$. The number with a left arrow indicates the number of outliers with even smaller probabilities that could not be represented in the plot.\label{hist_chisq_prob}}
\end{figure}

Figure~\ref{scatter_plot_best_sigma_mean_Fpeak} shows the distribution of $\sigma_\textrm{max}$ against $\overline{f}$. Note that a few objects have a relatively low significance as per the measure $\sigma_\textrm{max}$, but are still included as variable when above the threshold of the other indicator of variability $P(\chi^2,\nu)$. Figure~\ref{hist_chisq_prob} illustrates the distribution of outliers described by the parameter $P(\chi^2,\nu)$. The fraction of outliers to the right of the dashed line are those selected by the criterion $\sigma_\textrm{max}\ge 5$. The number with a left arrow indicates the number of outliers with even smaller probabilities that could not be represented in the plot.

\begin{figure} \centering
\includegraphics[width=\linewidth]{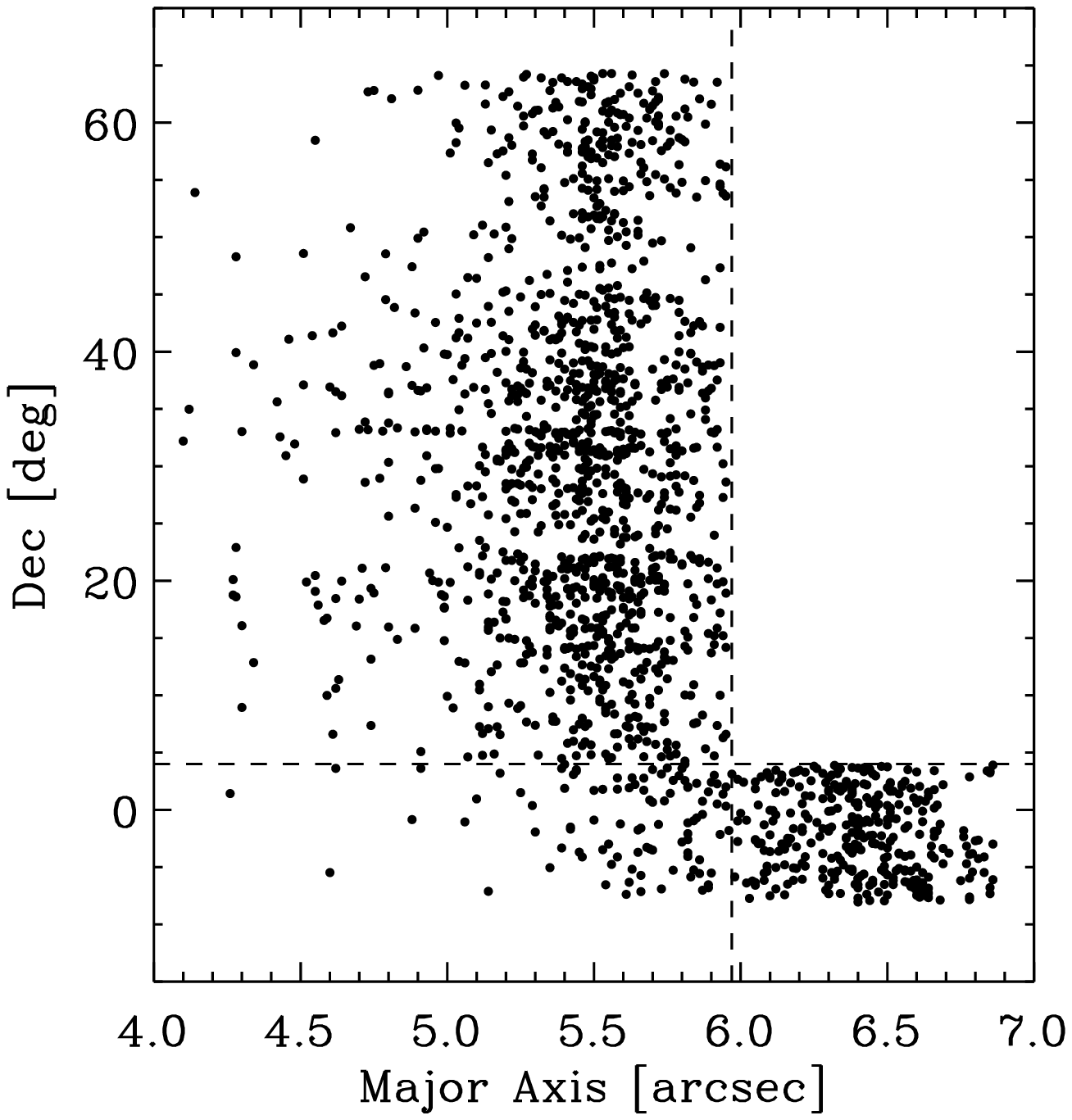}
\caption[Scatter plot of the sizes of outliers vs. the declination]{Scatter plot of the sizes of outliers vs. the declination. The horizontal dashed line shows the boundaries where the deconvolving beam size changes (cf. Figure~\ref{sky_positions}). The vertical dashed line arises from our definition of point-like -- a source with an intrinsic size $<$~2\farcs5.\label{dec_PS_size}}
\end{figure}

The sky positions of the outliers are shown in figure~\ref{sky_positions}. One notable feature is that the density of outliers below a declination of $\approx$~4\arcdeg is enhanced. This is because the synthesized beam with which the data is convolved increases below a declination of $\approx$~4\arcdeg as is clearly illustrated in figure~\ref{dec_PS_size}. Consequently, more objects meet the criterion for a point-like source, increasing the density of both non-variable and variable objects.  

\section{A Catalog of Highly Variable Radio Sources \label{catalog}}

Table~\ref{onepagesummarytable} presents a sample page from the list of 1627 highly variable radio sources identified in our study. The complete table is available online in electronic format. Column 1 provides the source position derived from the {\sl FIRST} catalog. Columns 2 and 3 give, respectively, the {\sl FIRST} catalog peak flux density and the mean peak flux density derived from the light curves. The range of flux densities for the source derived from the grid images are in column 4, while column 5 provides the NVSS catalog peak flux density. Note that, while {\sl FIRST} and NVSS have very different angular resolutions, our restriction of this study  to isolated point sources in the higher-resolution {\sl FIRST} survey means resolution effects should not be a factor in most cases.

These data are followed by the number of observations available (col. 6). The next three columns provide the three measures of variability: $\sigma_{P(\chi^2,\nu)}$ for the light curve (col. 7), the maximum deviation of a single data point from the mean flux density in the light curve (col. 8), and the difference (in $\sigma$) between the most significantly different pair of data points (col. 9). $\sigma_{P(\chi^2,\nu)}$ is the number of standard deviations on both sides of a normal distribution beyond which the probability from this normal distribution matches $P(\chi^2,\nu)$. The maximum-to-minimum flux density ratio is provided in column 10. The $T_\textrm{min}$~(days) in column 11 denotes the minimum timescale at which two data points differ by at least $6\sigma$. In cases where the maximum absolute difference of any pair of data points never reaches a value $6\sigma$, the value denotes the timescale at which the maximum absolute difference in column 10 is attained. In a few cases where the date of observation cannot be reliably obtained from the snapshot header, the entry is left without a numerical value. Column 12 contains a flag `N' if the outlier is found to be in the vicinity of a source brighter than 500~mJy within 31\arcmin. Column 13 provides the distance by which the mean normalized peak flux density of the outlier's neighbors departs from the mean of the distribution of such normalized neighbor peak flux densities from all outliers, and is a proxy for the quality of the field. The higher the absolute value in this column, the less ideal the quality of the field surrounding the outlier and hence, the less reliable the physical variability of the outlier.   

The type of light curve is denoted in column 14 by V (Variable) and T (Transient). The apparent SDSS-$i$~band magnitudes are provided in column 15 where a SDSS match is available. Finally, column 16 gives the best ranked counterpart at other wavelengths. This ranking is explained in \S\ref{crossid}.

\begin{figure} \centering
\includegraphics[width=\linewidth]{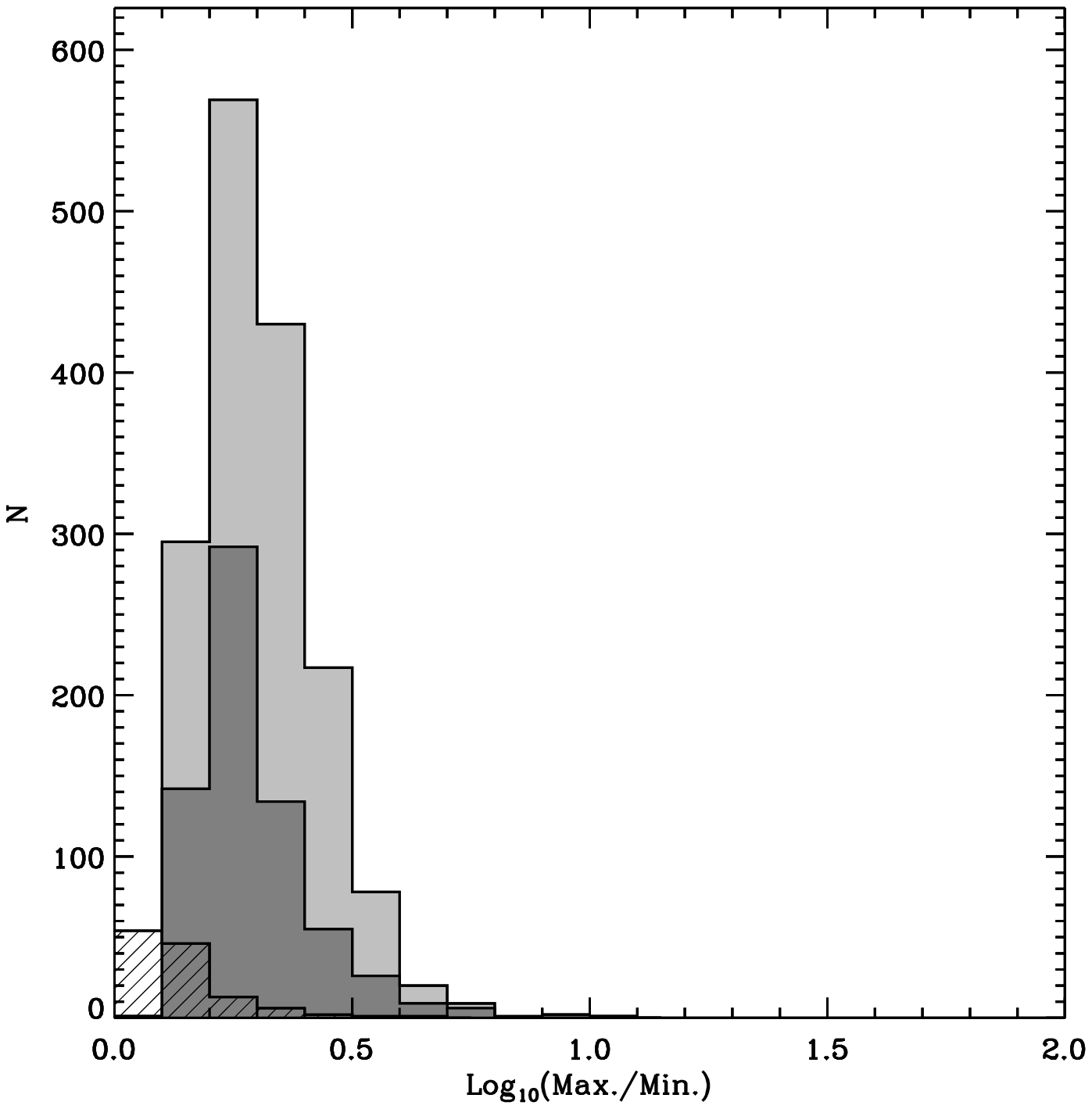}
\caption[Histogram of ratios of maximum to minimum peak flux densities of all variable objects]{Histogram of ratios of maximum to minimum peak flux densities of all of our variable objects. The hatched histogram is for objects from the sample of \citet{dev04}. The dark shading indicates the number of lower limits that fall inside this bin. These lower limits were derived using $3\sigma$ upper limits for non-detections in the denominator. \label{histogram_max_min_ratio_all}}
\end{figure}

Figure~\ref{histogram_max_min_ratio_all} shows the distribution of the ratio of maximum to minimum flux density for our 1627 variables which ranges from 25\% to a factor of $>$~20; overplotted is this ratio for the sample of variable sources \citep{dev04} with the same flux density distribution as ours. The areal density of their variables is $\sim$~1~deg$^{-2}$ while we find it to be $\sim$~0.2~deg$^{-2}$ for our sample. This difference is primarily due to the lower threshold of detection ($4\sigma$) used by \citet{dev04}. Sources for which lower limits are used are indicated by shading. The relation between maximum-to-minimum flux density ratio and mean peak flux density is shown as a scatter plot in figure~\ref{scatter_max_min_ratio_mean_Fpeak}.

\section{Identification of Variable Radio Sources \label{crossid}}

Table~\ref{crossID_summary_table} presents the results of cross-matching our catalog of variable sources with a variety of catalogs and databases. The first column lists the source type, followed by the specific catalogs used for cross identification and the matching radius adopted for each (col. 3). The rank in column 4 indicates the order in which the matches were applied; e.g., we first sought matches with bright star catalogs, then with a pulsar catalog, then with quasar catalogs, etc. This is important in interpreting the number of matches $N$ found in column 5 where we list both the number of total matches including those previously identified in a higher-ranked catalog and, in parentheses, the number of unique matches to the specified catalog. For example, the photometric SDSS quasar catalog has 68 total matches, but only 67 of them are real; the $68^\textrm{th}$ is also matched to a pulsar (which is the true identification).

Column 6 gives the match number as a percentage of all variable sources (e.g., a total of 7.4\% of all variables are SDSS quasars). Columns 7 and 8, respectively, give the percentage of all {\it FIRST} sources matching each catalog that are identified as variable (e.g., 76\% of all {\it FIRST}    
pulsars are variable, but only 1.2\% of SDSS spectroscopically identified quasars with {\it FIRST} counterparts are highly variable), and the percentage of all {\it FIRST} isolated point sources in the catalog that find identifications in the given catalog. Parentheses are used as in column 4.

In all cases, the rate of chance coincidences is low, typically less than one source per category with the exception of SDSS galaxies where up to seven chance coincidences (out of 442 matches) are possible. In summary, we find 5 bright stars, 13 radio pulsars, 120 quasars, and 489 galaxies among our variable sources, yielding firm identifications for 38\% of our objects. An additional 118 objects are identified with unclassified stellar objects in the SDSS and GSC~II databases; a total of 50 objects have X-ray counterparts, although only 5 of these are not also identified optically (mostly as quasars). This leaves 877 objects or 54\% of the total without identifications in other wavelength bands (although 499 of these are found in other radio catalogs). We discuss below each class of objects in turn, and compare the properties of the identified objects to the unidentified ones in order to seek clues as to the identities of the latter sources.

In figure~\ref{histogram_max_min_ratio_diff_types}, we show how the maximum-to-minimum flux density ratio differs for the five different classes of variable objects. The pulsars are, as expected the most highly variable, while the sources classified as galaxies have slightly higher variability amplitudes than the quasars. The objects with no counterparts closely mimic the distribution of galaxies.

\begin{figure} \centering
\includegraphics[width=\linewidth]{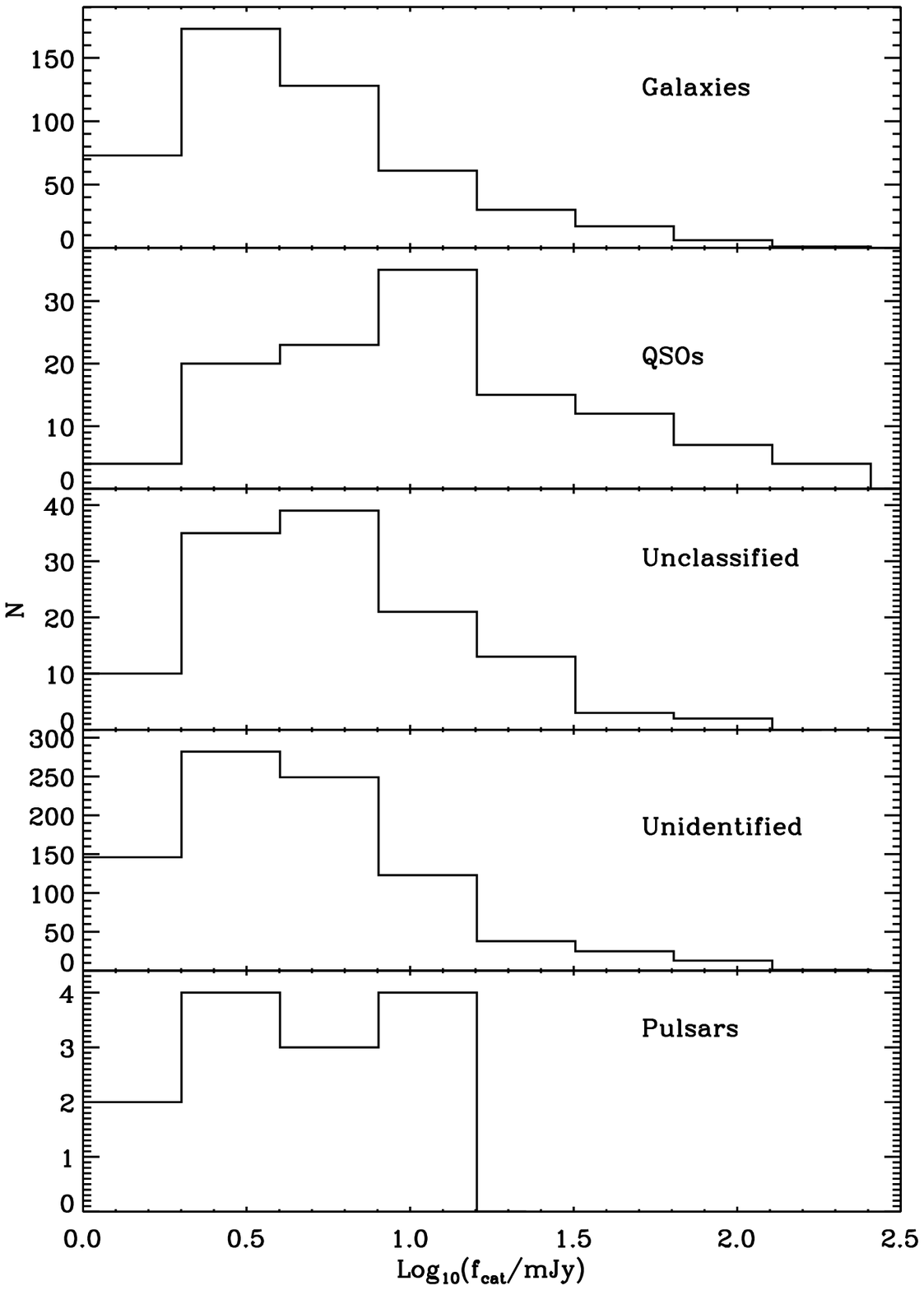}
\caption[Histogram of {\sl FIRST} catalog peak flux densities of different classes of variable objects]{Histogram of {\sl FIRST} catalog peak flux densities of different classes of variable objects. \label{histogram_radio_fluxes_diff_types}}
\end{figure}

Figure~\ref{histogram_radio_fluxes_diff_types} illustrates the distribution of mean peak flux densities for the various source classes. As expected, the relative fraction of QSOs with respect to galaxies increases with increasing mean peak flux density. The pulsars and the stars are much fainter with mean peak flux densities of a few mJy. Again, the distribution of unidentified objects appears to follow the trend shown by galaxies and interestingly, it also has a significant tail at the high-mean peak flux density end.

\subsection{Radio Pulsars}

While $\sim$~70 known radio pulsars fall within our survey area, most are both faint and have steep radio spectra, leading to only 17 detections in the {\sl FIRST} survey catalog (Table~\ref{PSR_summary_table}). Of these, 9 ($\sim$~50\%) appear in our catalog of highly variable sources. By slightly relaxing our criterion for a point-source, we find an additional four pulsars that would have been classified as variables by our algorithm; we add these to the total number of outliers quoted throughout the paper. Thus, more than three-quarters of the detected radio pulsars in the survey are found to be highly variable.

While the intrinsic radio luminosity of a pulsar is steady, its emission suffers from interstellar scintillation which arises when signals traveling in slightly different directions are alternately scattered into, and out of, our line of sight by the fluctuating electron density distribution of the interstellar medium. The effect is largest when the number of scattering centers is small, so the highest variability is seen for the nearest pulsars; indeed, 11 of our 13 variable objects have dispersion measures ($DM$) of $\lesssim$~20~cm$^{-3}$pc; the other two have $DM$ values of 27~cm$^{-3}$pc and 41~cm$^{-3}$pc. Of the four pulsars not seen to vary significantly, two have flux densities $<$~1.5~mJy, too close to our threshold to detect variability, and a third has a $DM>35$.

The typical time scale for scintillation is $\sim$~1 min at 20 cm (to be compared with our 165 s integration time) and the decorrelation bandwidth is typically smaller than our 50~MHz bandwidth, so our observational parameters largely smooth over the fluctuations. Nonetheless, pulsars are among the most variable objects in our sample, with 8 out of 13 varying by more than a factor of 4 (see figure~\ref{histogram_max_min_ratio_all}). Interestingly, 17 of the 877 variables which lack optical counterparts entirely also vary by this large factor; this is to be contrasted to the fact that none of the 120 sources identified as quasars are this variable, and only 1\% of the sources coincident with galaxies vary by such large factors. All these 17 sources have flux densities less than 4~mJy and most appear only once or twice, falling below our detection threshold in the other observations. An examination of the radio spectrum and polarization of these sources could provide candidates for pole-on, very short period, and/or intermittent pulsars worthy of further study.

\subsection{Optical Counterparts}

\begin{figure} \centering
\includegraphics[width=\linewidth]{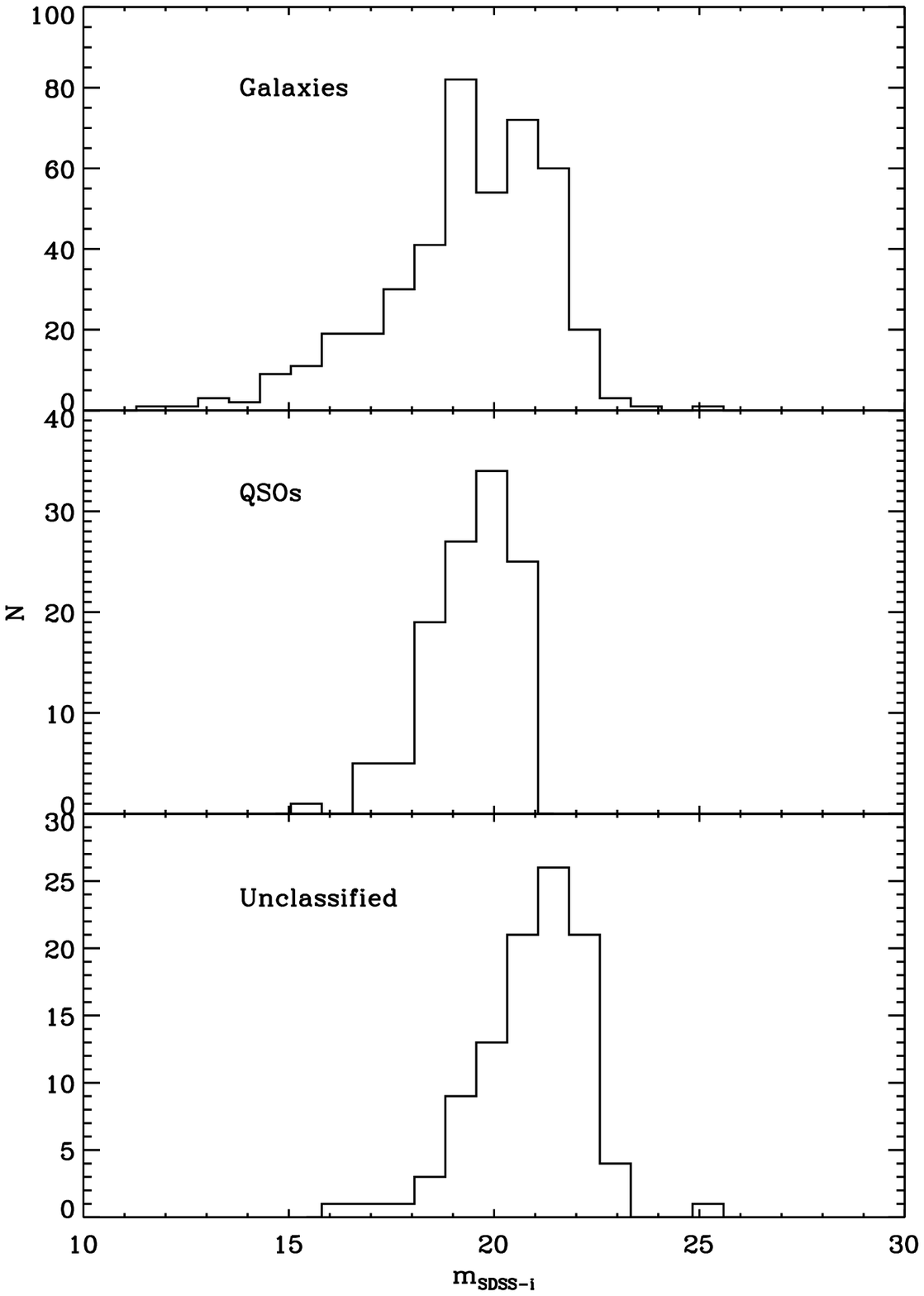}
\caption[Histogram of SDSS {\it i}-band magnitudes of different classes of variable objects detected in SDSS]{Histogram of SDSS {\it i}-band magnitudes of different classes of variable objects detected in SDSS. \label{histogram_magnitudes_diff_types}}
\end{figure}

A total of 732 radio variables have optical counterparts in one or more of the catalogs used in the identification program summarized in table~\ref{crossID_summary_table}; the vast majority of these are from a match to the SDSS DR7 catalog.
Figures~\ref{histogram_magnitudes_diff_types} shows SDSS {\it i-}band magnitudes for the various source classes, while figure~\ref{scatter_plot_colors_diff_types_1} displays SDSS color-color diagrams for the identified galaxies and quasars of our sample.  We discuss each sample of objects in turn below. 

\subsubsection{Stars}

Most stars are very faint at radio wavelengths. Of the more than 800,000 sources in the {\sl FIRST} catalog, only 37 match objects in the bright star catalogs we examined to within 1\farcs4. To check that the potentially  high proper motions of these (mostly) nearby stars have not obviated any true matches, we expanded the search radius to 5\arcsec; no additional matches were found when proper-motion-corrected positions were used. \citet{hel99} conducted a more exhaustive search of the first 4,760~deg$^2$ of the {\sl FIRST} survey and found a total of 26 radio-detected stars, 5 of which were at flux densities below the survey limit of 1.0~mJy. One of the detections reported in that earlier survey, the flare star (08 08 55.47 +32 49 06.0), is found in our catalog of highly variable objects. It has a light curve showing variability on timescales of days, months and years. In total, we find only 5 stars both bright enough and variable enough to make our sample of variables and transients. All are previously known variables.

\subsubsection{Quasars}

A total of 53 quasars from the SDSS DR5 spectroscopic quasar catalog \citep{sch07} are found in our list of variables. In addition, another 67 objects in the DR6 photometric quasar catalog of \citet{ric09} are in our list, making a total of 120 quasars in the sample or 7\% of all variables. For the whole {\sl FIRST} catalog there are 8396 and 10,025 matches from the spectroscopic and photometric catalogs, respectively, accounting for nearly 2.5\% of all {\sl FIRST} sources. Table~\ref{crossID_summary_table}, however, includes only matches with isolated point-like sources from the {\sl FIRST} catalog.

The quasars have the highest mean radio flux density ($>10$~mJy) among our identified source classes (figure~\ref{histogram_radio_fluxes_diff_types}) and the lowest mean max-to-min flux density ratio. This is in part a selection effect, in that only in bright sources does modest variability exceed our significance threshold. Nonetheless, with our coarse and uneven sampling, we have found only a dozen quasars that vary by more than a factor of 2.5.

\subsubsection{Galaxies}

The largest identified segment of our sample is coincident with objects classified as galaxies in SDSS (with a few additional galaxies contributed from GSC~II, 2MASS, NED, etc.). The 489 galaxies comprise roughly 30\% of our sample. The galaxies have a broad spread in magnitudes (figure~\ref{histogram_magnitudes_diff_types}) and  a lower mean radio flux density than the quasars but a somewhat higher amplitude of variability (Figure~\ref{histogram_max_min_ratio_all}).

Since our matching radius is only 1\farcs4, the radio sources are all coincident, within the uncertainties, with the galaxy nuclei. AGN are thus undoubtedly the source of variable radio emission in virtually all of these objects.
Optical spectroscopy (and/or hard X-ray imaging) of this sample could be interesting, in that it could reveal the fraction of buried AGN which show no optical evidence of an accreting black hole.

We examined by eye SDSS images for all 1627 variable source candidates. Virtually all of the matched sources had clear counterparts and almost all of the unmatched sources were blank fields as expected. However, there were a handful of cases where there was a coincident optical object that did not appear in the SDSS catalog; these are noted in the table. In addition, there were several cases in which the radio source was superposed on a large ($>$10\arcsec) galaxy and not coincident with the galaxy nucleus. In a few cases, a backgound source was visible through the galaxy light and is the likely radio source counterpart. In several other cases, however, no background source is apparent although it could, of course, simply be obscured by the foreground galaxy's light or dust. 

\subsubsection{Unclassified Stellar Counterparts}

A total of 114 radio variables are coincident with unclassified stellar objects in the SDSS DR7 catalog; a few additional such objects are found in the GSC~II catalog in areas lying outside of the SDSS coverage. Figure~\ref{scatter_plot_colors_diff_types_2} shows the SDSS color-color diagrams for these stellar objects; comparison with figure~\ref{scatter_plot_colors_diff_types_1} shows a large majority of these objects have colors consistent with quasars, while virtually all of the remainder fall within the galaxy contours in color-color space. The majority of these objects fall below the $m_i=21.0$ cutoff for the SDSS photometric quasar catalog, but nearly 60 objects lie above this threshold. Spectroscopic follow-up of these objects could provide insight into the completeness of the photometric quasar catalog\footnote{The three brightest objects with $m_i<18$ include one source with no spectrum and two with spectra classified as `STAR'.}.
\subsubsection{Unidentified Sources}

There are no optical (or X-ray) counterparts for 877 sources comprising 54\% of our sample. Figures~\ref{histogram_max_min_ratio_all} and \ref{scatter_plot_colors_diff_types_1} show that these sources mostly cohabit the parameter-space of galaxies and QSOs, although there are a few notable outliers with very high variability amplitudes as noted above.

\section{Summary}\label{summary}

After analyzing $\sim 55,000$ snapshot images from the {\sl FIRST} survey of the radio sky covering 8444 deg$^2$ with sensitivity down to 1 mJy for variability and transient phenomena, we have assembled a sample of 1627 sources that are significantly variable. This sample was matched with multi-wavelength catalogs to identify counterparts. We found 13 radio pulsars, 53 SDSS spectroscopic QSOs, 67 SDSS photometric QSOs, 489 galaxies, 5 stars, 123 optically detected but unclassified sources and 877 objects lacking optical counterparts. The unidentified sources mostly occupy the parameter-space of galaxies and QSOs but there are a few notable outliers. Follow-up observations of several of these source classes would likely be fruitful.

Our study has shown that there is much to be discovered in the dynamic radio sky. Exploration by the next-generation instruments sampling different portions of the spectral and temporal domains will prove to be highly productive.   

\acknowledgments We are grateful to Jacqueline van Gorkom and Zoltan Haiman for their valuable comments on this paper. RHB's work was supported in part under the auspices of the US Department of Energy by Lawrence Livermore National Laboratory under contract W-7405-ENG-48.

The National Radio Astronomy Observatory is a facility of the National Science Foundation operated under cooperative agreement by Associated Universities, Inc. This research has made use of the NASA/IPAC Extragalactic Database (NED) which is operated by the Jet Propulsion Laboratory, California Institute of Technology, under contract with the National Aeronautics and Space Administration. 

Funding for the SDSS and SDSS-II has been provided by the Alfred P. Sloan Foundation, the Participating Institutions, the National Science Foundation, the U.S. Department of Energy, the National Aeronautics and Space Administration, the Japanese Monbukagakusho, the Max Planck Society, and the Higher Education Funding Council for England. The SDSS Web Site is \url{http://www.sdss.org/}.

The SDSS is managed by the Astrophysical Research Consortium for the Participating Institutions. The Participating Institutions are the American Museum of Natural History, Astrophysical Institute Potsdam, University of Basel, University of Cambridge, Case Western Reserve University, University of Chicago, Drexel University, Fermilab, the Institute for Advanced Study, the Japan Participation Group, Johns Hopkins University, the Joint Institute for Nuclear Astrophysics, the Kavli Institute for Particle Astrophysics and Cosmology, the Korean Scientist Group, the Chinese Academy of Sciences (LAMOST), Los Alamos National Laboratory, the Max-Planck-Institute for Astronomy (MPIA), the Max-Planck-Institute for Astrophysics (MPA), New Mexico State University, Ohio State University, University of Pittsburgh, University of Portsmouth, Princeton University, the United States Naval Observatory, and the University of Washington.

This research has made use of data obtained from the High Energy Astrophysics Science Archive Research Center (HEASARC), provided by NASA's Goddard Space Flight Center.

\clearpage

\clearpage

\clearpage

\clearpage

\begin{FPfigure} \centering
\includegraphics[width=\linewidth]{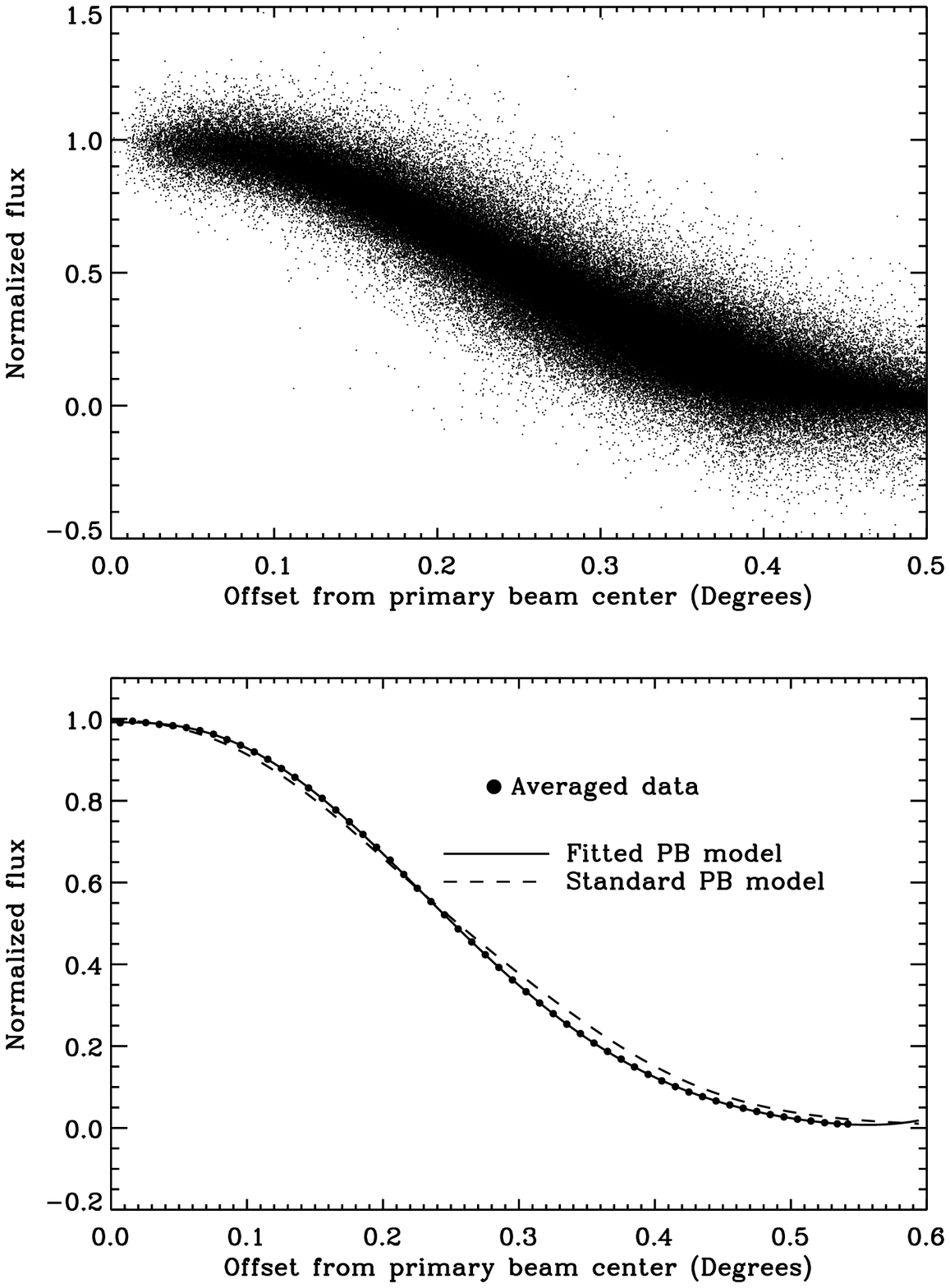}
\caption[Empirical primary beam pattern]{\textit{Upper panel}: Fitted peak Gaussian flux densities of isolated point sources in the snapshot images normalized by their respective catalog flux densities. \textit{Lower panel:} Binned average (filled circles) of the panel above over-plotted with the fitted polynomial primary beam model (solid line) and the nominal primary beam correction factor found in AIPS (`plus' symbols).\label{PBEAM}}
\end{FPfigure}

\clearpage

\begin{FPfigure} \centering
\includegraphics[width=\linewidth]{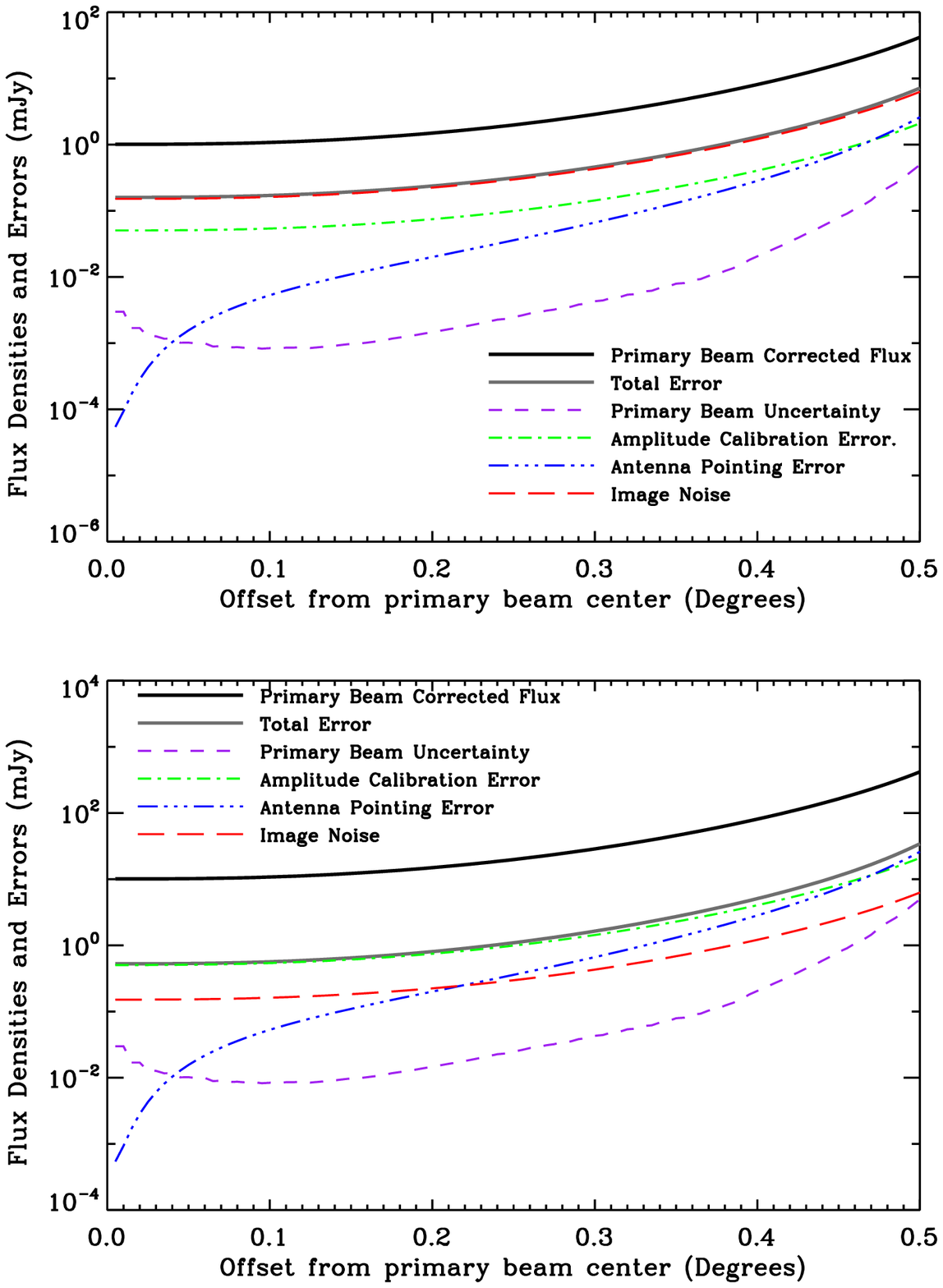}
\caption[Relative contributions of errors from various causes]{Relative contributions of errors from various causes. 
  \textit{Upper panel}: For a source with an observed peak flux density of 1~mJy. \textit{Lower panel}: For a source with an observed peak flux density of 10~mJy. The solid line shows the primary beam-corrected intrinsic flux densities of sources as a function of off-axis angle.\label{errors}}
\end{FPfigure}

\clearpage

\clearpage

\clearpage

\clearpage

\clearpage

\clearpage

\clearpage

\begin{FPfigure} \centering
\includegraphics[width=\linewidth]{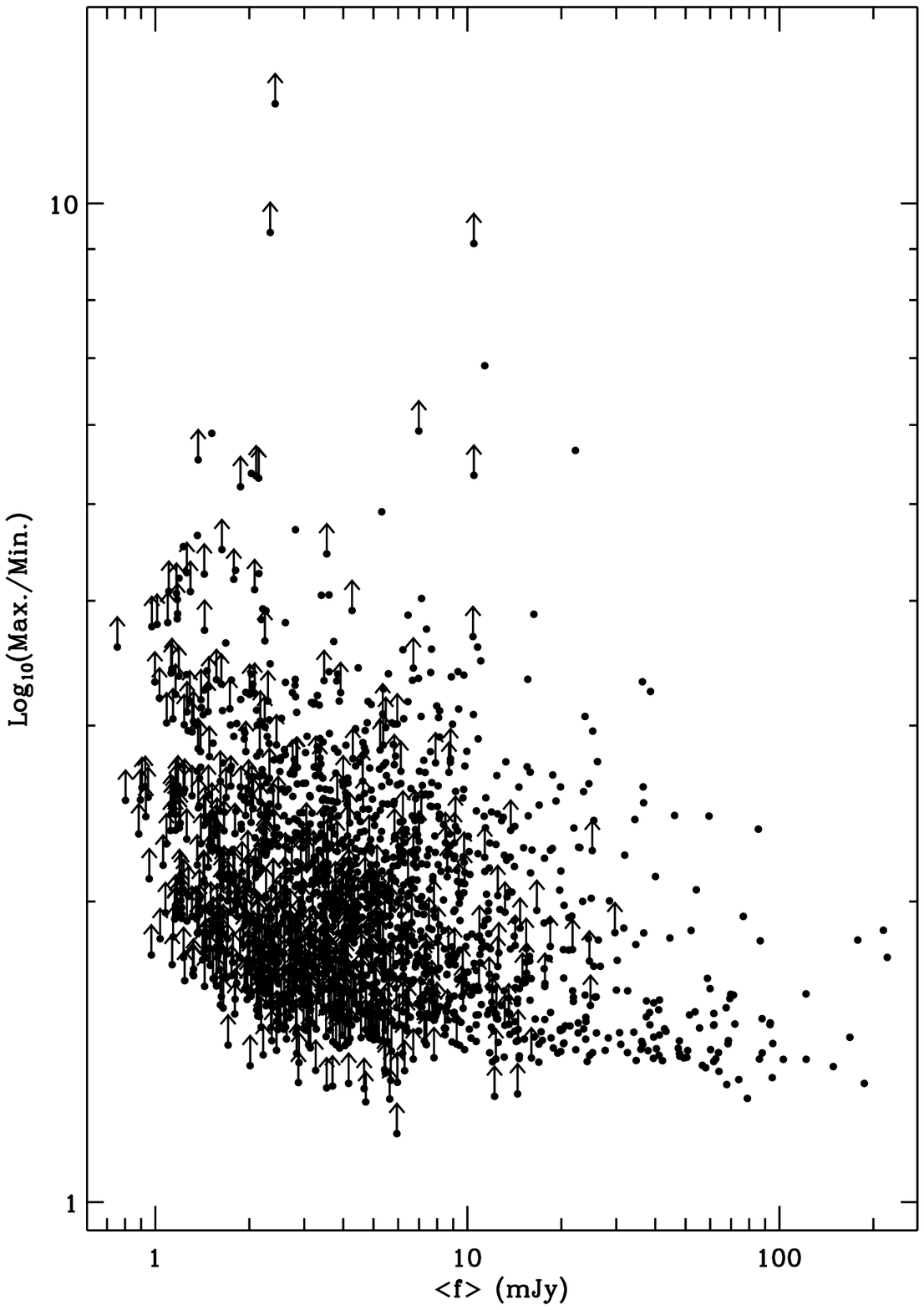}
\caption[Scatter plot of the ratios of maximum to minimum peak flux densities of all objects vs. their average peak flux densities]{Scatter plot of the ratios of maximum to minimum peak flux densities of all objects against their average peak flux densities. The upward arrows denote that the value of the ratio is derived from a minimum peak flux density which was a non-detection whose $3\sigma$ upper limit was used, correspondingly yielding a lower limit for the ratio. The numerator-denominator pair in the ratio comes from the pair of data points in the light curve that yields the maximum absolute value of $\Delta_\textrm{max}$. \label{scatter_max_min_ratio_mean_Fpeak}}
\end{FPfigure}

\clearpage

\begin{FPfigure} \centering
\includegraphics[width=\linewidth]{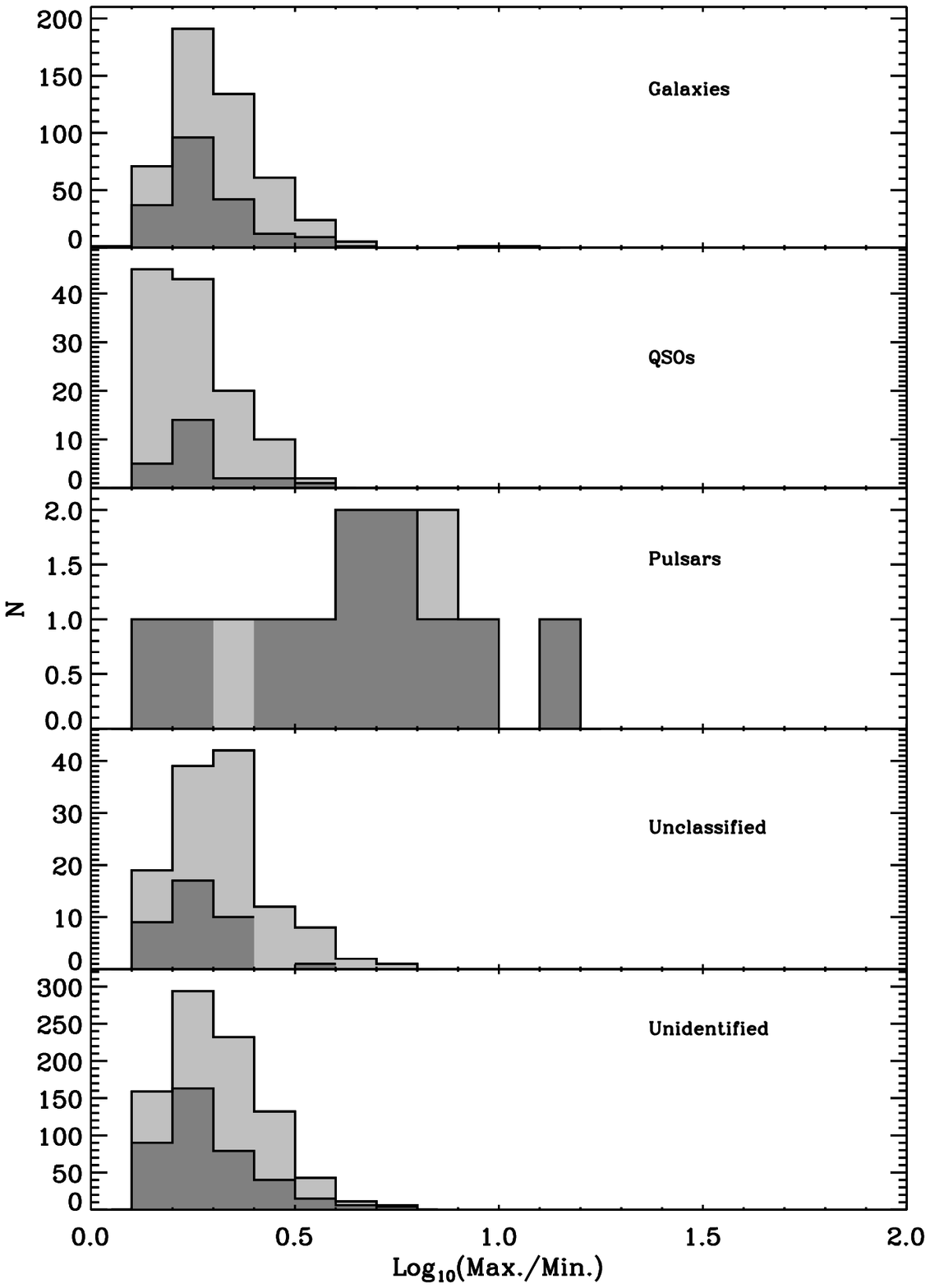}
\caption[Histogram of ratios of maximum to minimum peak flux densities of different classes of variable objects]{Histogram of ratios of maximum to minimum peak flux densities of different classes of variable objects. The dark shading indicates the number of lower limits that fall inside this bin. These lower limits were derived using $3\sigma$ upper limits for non-detections in the denominator. \label{histogram_max_min_ratio_diff_types}}
\end{FPfigure}

\clearpage

\clearpage

\clearpage

\begin{FPfigure} \centering
\includegraphics[width=\linewidth]{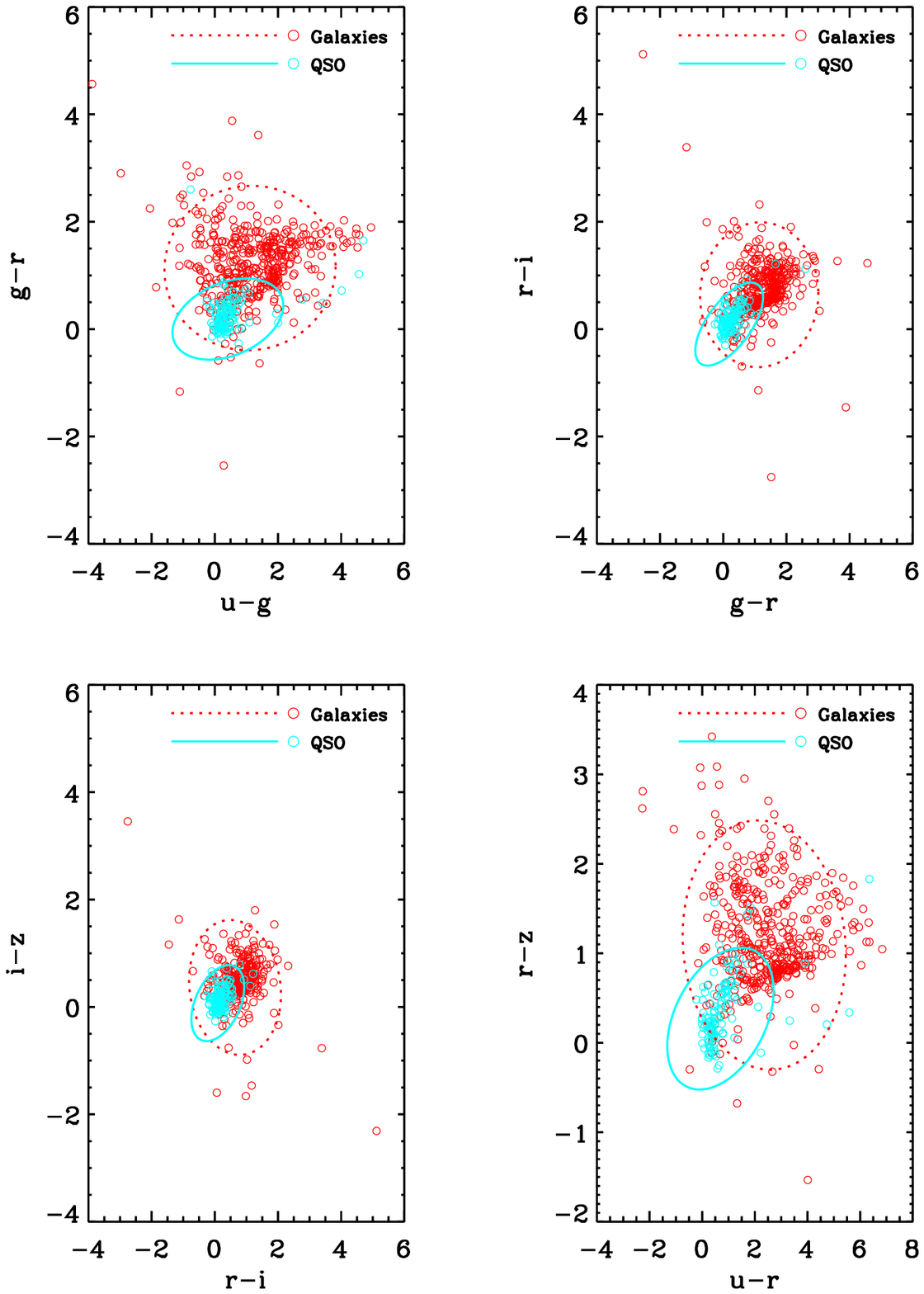}
\caption[Scatter plot of SDSS colors of  variable objects detected and classified as galaxies and QSOs in SDSS]{SDSS color-color diagrams for  variable objects detected and classified as galaxies and QSOs in SDSS. The red and green ellipses denote the color-color distribution of galaxies and QSOs respectively.\label{scatter_plot_colors_diff_types_1}}
\end{FPfigure}

\clearpage

\begin{FPfigure} \centering
\includegraphics[width=\linewidth]{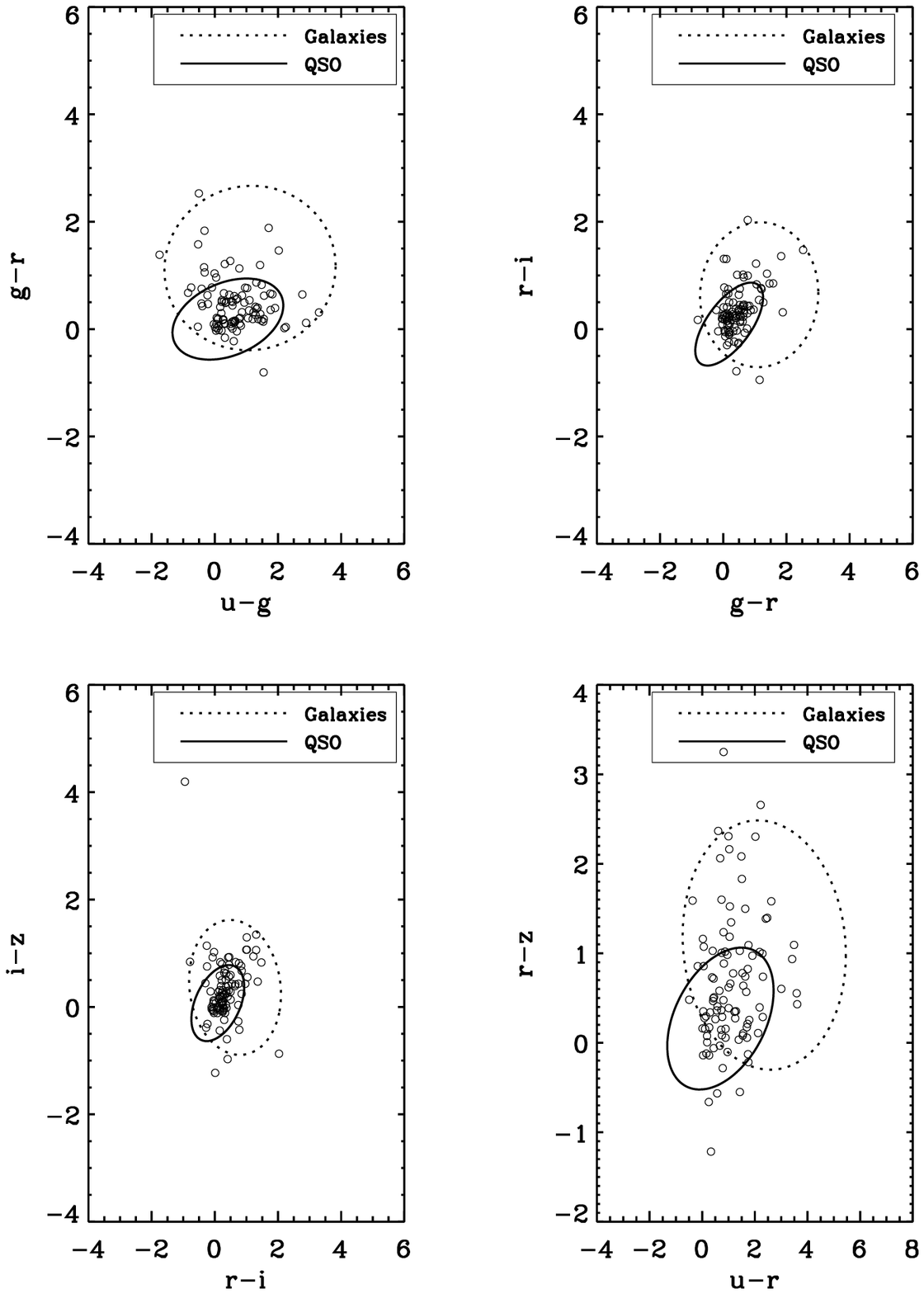}
\caption[Scatter plot of SDSS colors of  variable objects detected but unclassified in SDSS]{Scatter plot of SDSS colors of  variable objects detected but unclassified in SDSS. The dotted and dashed ellipses denote the color-color distribution of galaxies and QSOs respectively, identical to those in figure~\ref{scatter_plot_colors_diff_types_1}. The open circles in the legend represent the numbers in the plot indicating the unclassified objects that fall outside of both ellipses. The numbering scheme is used for easy identification of objects that consistently fall outside of the QSO and galaxy ellipses. \label{scatter_plot_colors_diff_types_2}}
\end{FPfigure}

\clearpage

\begin{deluxetable}{lrrcrrrrrcccrccl}
\tabletypesize{\tiny}
\rotate
\tablecolumns{16}
\tablewidth{0pt}
\tablecaption{Summary of Properties of Variables \& Transients \label{onepagesummarytable}}
\tablehead{ \colhead{Coordinates (J2000)}    &  \multicolumn{4}{c}{$f$ (mJy)} &  \colhead{N}   &  \multicolumn{3}{c}{Measures} & \colhead{$\frac{\textrm{Max.}}{\textrm{Min.}}$} & \colhead{$T_\textrm{min}$} & \colhead{Flags} & \colhead{$\sigma_\textrm{nbr}$} & \colhead{Type}   &  \colhead{Mag.}  &   \colhead{Cross-ID} \\
\cline{2-5} \cline{7-9}  \\
\colhead{} & \colhead{Cat.} & \colhead{$\overline{f}$}   & \colhead{Range} & \colhead{NVSS} & \colhead{} & \colhead{$\sigma_{P(\chi^2,\nu)}$} & \colhead{$\sigma_\textrm{max}$} & \colhead{$\Delta_\textrm{max} (\sigma)$}   & \colhead{} & \colhead{(days)} & \colhead{} & \colhead{} & \colhead{}  & \colhead{i}  & \colhead{} }
\startdata
 12 01 05.881  +20 23 10.85 & 5.91 & 5.85 & 2.56 - 6.19 & 4.28 & 3 &       4.86 & 5.18 & 4.78 & 2.33 & 5 & \nodata & 0.09 & V & 21.24 & SDSS-G \\
 12 01 07.863  +24 53 33.32 & 8.78 & 8.96 & 3.04 - 9.15 & 8.08 & 3 &       6.97 & 7.27 & 6.50 & 3.01 & 2 & \nodata & -0.75 & V & \nodata & \nodata \\
 12 01 35.629  +33 12 33.28 & 4.75 & 4.53 & $<$3.05 - $<$9.02 & 4.00 & 6 &       5.70 & 5.61 & 5.35 & $>$1.58 & \nodata & \nodata & -1.53 & V & \nodata & \nodata \\
 12 01 49.193  +03 39 03.44 & 9.18 & 9.05 & 4.67 - 9.58 & 8.73 & 3 &       4.90 & 5.22 & 4.64 & 1.95 & 1 & \nodata & 0.19 & V & 19.73 & SDSS-G \\
 12 02 10.433  +16 33 28.02 & 1.25 & 1.28 & $<$1.05 - $<$3.75 & \nodata & 4 &       5.32 & 4.06 & 5.25 & $>$2.02 & 1 & \nodata & 0.23 & V & 21.28 & SDSS-G \\
 12 02 25.829  +30 37 38.12 & 4.44 & 4.46 & $<$2.47 - $<$7.81 & 3.76 & 3 &       6.51 & 6.39 & 6.17 & $>$1.97 & 3 & \nodata & 0.65 & V & 19.39 & SDSS-G \\
 12 02 40.649  +21 56 20.99 & 8.40 & 7.14 & 6.76 - 11.24 & 9.19 & 3 &       5.27 & 5.27 & 5.28 & 1.66 & 0 & \nodata & -3.17 & V & 20.14 & SDSS-G \\
 12 02 49.673  +31 36 40.33 & 2.51 & 2.70 & $<$1.67 - $<$15.62 & \nodata & 6 &       6.10 & 5.48 & 5.70 & $>$1.83 & 95 & \nodata & -0.95 & V & \nodata & \nodata \\
 12 02 53.621  +34 31 29.70 & 4.54 & 4.47 & 1.43 - $<$15.83 & 4.61 & 5 & $\geq$8.00 & 7.09 & 6.79 & 3.43 & 5 & \nodata & 0.68 & V & \nodata & \nodata \\
 12 02 55.734  +04 33 47.25 & 2.23 & 2.21 & 1.39 - 3.10 & \nodata & 3 &       5.19 & 3.65 & 4.87 & $>$2.24 & 388 & N & -0.26 & V & \nodata & \nodata \\
 12 03 02.267  +03 37 53.99 & 2.49 & 2.60 & $<$1.66 - 3.29 & \nodata & 4 &       6.61 & 5.99 & 6.56 & $>$1.99 & 1 & \nodata & -0.57 & V & 19.26 & SDSS-QSO(P) \\
 12 03 19.427  +04 38 15.57 & 3.64 & 3.58 & 2.24 - $<$13.88 & 3.39 & 4 &       6.36 & 5.13 & 5.72 & $>$1.73 & 388 & N & -0.12 & V & \nodata & \nodata \\
 12 03 57.217  -01 38 26.51 & 2.20 & 2.34 & $<$1.73 - $<$7.93 & \nodata & 4 &       5.51 & 5.28 & 5.61 & $>$1.70 & 1 & \nodata & 1.20 & V & 16.02 & SDSS-G \\
 12 04 28.315  +03 52 59.25 & 5.38 & 5.42 & $<$4.02 - 7.42 & \nodata & 4 &       5.70 & 4.92 & 5.35 & $>$1.53 & 1 & N & 0.62 & V & \nodata & \nodata \\
 12 04 39.877  +50 28 20.71 & 2.32 & 2.32 & 1.43 - 3.59 & \nodata & 3 &       5.98 & 4.80 & 5.94 & 2.50 & 27 & \nodata & -0.10 & V & \nodata & \nodata \\
 12 04 50.878  +29 20 33.40 & 1.90 & 1.66 & $<$1.40 - 2.96 & \nodata & 3 &       5.63 & 4.35 & 5.92 & $>$2.12 & 11 & \nodata & -1.05 & V & 21.79 & SDSS-G \\
 12 04 53.441  -04 59 41.50 & 4.62 & 4.52 & 1.71 - 4.70 & \nodata & 3 &       6.30 & 6.49 & 5.94 & 2.76 & 40 & \nodata & -1.43 & V & \nodata & \nodata \\
 12 05 26.547  +32 54 33.42 & 3.53 & 3.39 & 1.80 - 4.03 & 4.21 & 4 &       5.11 & 5.32 & 4.98 & 2.24 & 488 & \nodata & 0.44 & V & 15.29 & SDSS-G \\
 12 05 51.716  +21 11 20.45 & 10.99 & 11.12 & 6.29 - 13.44 & 8.62 & 5 &       6.11 & 6.53 & 5.50 & 1.82 & 5 & \nodata & 0.23 & V & \nodata & \nodata \\
 12 06 06.530  +61 42 04.63 & 23.40 & 25.19 & 24.94 - 36.24 & 25.18 & 5 &       5.43 & 3.92 & 3.61 & 1.41 & 0 & \nodata & 0.18 & V & \nodata & \nodata \\
 12 06 08.024  +36 43 31.29 & 3.11 & 3.01 & $<$2.06 - $<$24.94 & \nodata & 6 &       5.28 & 5.33 & 5.40 & $>$1.62 & 5 & \nodata & -0.41 & V & 21.89 & SDSS-G \\
 12 06 18.560  -07 36 45.19 & 6.53 & 7.24 & 4.32 - 8.39 & 9.55 & 2 &       7.32 & 6.93 & 6.31 & 1.94 & 0 & N & -0.25 & V & \nodata & \nodata \\
 12 06 26.572  +00 09 31.41 & 4.44 & 4.35 & 2.60 - 10.80 & 4.47 & 4 &       5.18 & 4.98 & 4.75 & 1.86 & 3 & \nodata & 0.93 & V & 18.13 & SDSS-G \\
 12 07 28.795  +25 30 18.66 & 17.03 & 16.69 & 16.41 - 38.70 & 13.66 & 6 &       5.40 & 4.93 & 5.18 & $>$1.96 & 0 & \nodata & -0.06 & V & 19.62 & SDSS-G \\
 12 07 47.271  -06 08 57.51 & 4.11 & 4.08 & 1.87 - $<$13.68 & \nodata & 4 &       6.84 & 6.63 & 5.92 & 2.87 & 2 & \nodata & -0.96 & V & \nodata & \nodata \\
 12 08 06.423  +36 53 12.45 & 2.84 & 2.75 & 1.65 - 5.35 & 2.66 & 3 &       5.31 & 4.85 & 5.64 & 3.24 & 6 & \nodata & 0.10 & V & \nodata & \nodata \\
 12 08 12.253  +33 13 22.10 & 2.37 & 2.61 & 1.50 - 4.22 & 3.29 & 4 &       6.95 & 5.35 & 6.76 & 2.81 & 483 & \nodata & 0.35 & V & \nodata & \nodata \\
\enddata

\end{deluxetable}

\clearpage

\begin{deluxetable}{llcccccc} 
\tabletypesize{\tiny} 
\tablewidth{0pt}
\tablecaption{Summary of Cross-ID Statistics for Variables \& Transients.\label{crossID_summary_table}}

\tablehead{
\colhead{Cross-ID} & \colhead{Catalog} & \colhead{Rank\tablenotemark{a}} & \colhead{$\Delta\theta$ (\arcsec)} & \multicolumn{2}{c}{Matches} & \multicolumn{2}{c}{Match Rate} \\
\cline{5-6} \cline{7-8}  \\
\colhead{} & \colhead{} & \colhead{} & \colhead{} & \colhead{N} & \colhead{\% of Variables} & 
\colhead{Variable \%} & \colhead{\% of {\sl FIRST} sources\tablenotemark{b}}
}

\startdata

                           \multirow{2}{*}{STAR}      &                          HIPPARCOS                    &  1  &   1.4 & 3 (3)     &       0.18 (0.18) &   27.3 (27.3)    &       0.004 (0.004) \\
                                                      &                          TYCHO                        &  2  &   1.4 & 4 (2)     &       0.25 (0.12) &   22.2 (22.2)    &       0.006 (0.003) \\
                           
\cline{1-8} \\					              	        	 
                           PULSAR\tablenotemark{c}    &                          ATNF                         &  3  &   3.0 & 13 (13)   &       0.8 (0.8) &   76.48 (76.48)  &      0.006 (0.006) \\      
\cline{1-8} \\						      	        	 
                           \multirow{2}{*}{QSO}       &                          SDSS (S)\tablenotemark{d}    &  4  &   1.4 & 53 (53)   &       3.26 (3.26) &   1.16 (1.16)    & 1.632 (1.632)      \\
                                                      &                          SDSS (P)\tablenotemark{e}    &  5  &   1.4 & 68 (67)   &       4.18 (4.12) &   1.1 (1.09)     & 2.203 (2.239)      \\
\cline{1-8} \\						      	        	 
                           \multirow{4}{*}{GALAXY}    &                          SDSS DR7                     &  6  &  1.4  & 442 (431) &      27.17 (26.49) &   0.55 (0.54)    &  28.92 (29.68)      \\
                                                      &                          GSC-2\tablenotemark{f}       &  7  &  1.4  & 349 (56)  &      21.45 (3.44)  & \tablenotemark{f}  & \tablenotemark{f} \\
                                                      &                          2MASS\tablenotemark{g}       &  8  &  1.4  & 121 (1)   &      7.44 (0.06)   & \tablenotemark{g}  & \tablenotemark{g} \\ 
                                                      &                          APMUKS(BJ)\tablenotemark{h}  & 13  &  6.0  & 2 (1)     &      0.12 (0.06)   & \nodata                & \nodata               \\
\cline{1-8} \\						      	        	 
              \multirow{7}{*}{UNCLASSIFIED}           &                          SDSS DR7                     &  6  &  1.4  &  227 (114) &     13.95 (7.01) &  0.91 (0.75)        & 8.9 (5.7)         \\
                                                      &                          GSC-2\tablenotemark{f}       &  7  &  1.4  &  111 (4)   &      6.82 (0.25)  & \tablenotemark{f}  & \tablenotemark{f} \\
                                                      &                          2MASS\tablenotemark{g}       &  8  &  1.4  &  27  (0)   &      1.66 (0.00)  & \tablenotemark{g}  & \tablenotemark{g} \\ 
                                                      &                          CHANDRA                      &  9  &  3.0  &  7   (1)   &      0.43 (0.06)  & 1.45 (1.23)        & 0.173 (0.048)     \\
                                                      &                          XMM                          & 10  & 15.0  &  15  (2)   &      0.92 (0.12)  & 1.52 (1.03)        & 0.353 (0.116)     \\   
                                                      &                          RASS-BSC                     & 11  & 30.0  &  15  (0)   &      0.92 (0.00)  & 2.29 (0.00)        & 0.234 (0.011)     \\ 
                                                      &                          RASS-FSC                     & 12  & 60.0  &  13  (2)   &      0.8  (0.12)  & 0.70 (0.49)        & 0.664 (0.246)     \\
\cline{1-8} \\			
                        \multirow{8}{*}{UNIDENTIFIED} &                          GB-87\tablenotemark{h}       & 13  &  6.0                 & 21  (7)   &   1.29 (0.43)  & \nodata         & \nodata                      \\
                                                      &                           VLSS\tablenotemark{h}       & 13  &  6.0                 & 5   (2)   &   0.31 (0.12)  & \nodata         & \nodata                      \\
                                                      &                          WB-92\tablenotemark{h}       & 13  &  6.0                 & 13  (1)   &   0.8  (0.06)  & \nodata         & \nodata                      \\
                                                      &                       ABELL-04\tablenotemark{h}       & 13  &  6.0                 & 1   (1)   &   0.06 (0.06)  & \nodata         & \nodata                      \\
                                                      &                             B3\tablenotemark{h}       & 13  &  6.0                 & 1   (1)   &   0.06 (0.06)  & \nodata         & \nodata                      \\
                                                      &                             WN\tablenotemark{h}       & 13  &  6.0                 & 2   (1)   &   0.12 (0.06)  & \nodata         & \nodata                      \\
                                                      &                           NVSS                        & 14  &  7.1                 & 989 (487) &  60.79 (29.93) & 0.75 (0.62) & 47.46 (46.88)            \\                    
                                                      &                           NONE                        & 15  & \nodata\tablenotemark{i} & 377 (377) & 23.17 (23.17)     & 0.43        & \nodata (100.0)            \\  
			                                                      
\cline{1-8} \\			
{\bf TOTAL} & & & & {\bf 1623+4\tablenotemark{c}} & {\bf 100.0} & & \\
\enddata
\tablenotetext{a}{Rank indicates the order in which the matches are made with different catalogs.}
\tablenotetext{b}{Includes only the isolated point-like sources from the {\sl FIRST} catalog}
\tablenotetext{c}{Four pulsars are slightly resolved but nevertheless included; more detailed data available in table \ref{PSR_summary_table}}
\tablenotetext{d}{Data obtained from SDSS DR5 spectroscopic QSO sample}
\tablenotetext{e}{Data obtained from SDSS DR6 photometric QSO sample}
\tablenotetext{f}{Has solely made use of the Guide Star Catalog II cross-match information available in the {\sl FIRST} catalog}
\tablenotetext{g}{Has solely made use of the 2 Micron All-Sky Survey cross-match information available in the {\sl FIRST} catalog}
\tablenotetext{h}{Data obtained from the NASA Extragalactic Database}
\tablenotetext{i}{Search radius varies depending on the catalog being searched}


\end{deluxetable}

\clearpage

\begin{deluxetable}{rrlrrrr} 
\tabletypesize{\scriptsize} 
\tablecaption{Summary of the properties of pulsars.\tablenotemark{a} \label{PSR_summary_table}}
\tablewidth{0pt}
\tablehead{
\colhead{Coordinates (J2000)} & \colhead{$\Delta$ (\arcsec)} & \colhead{PSR Name} & \colhead{$f_\textrm{cat}$ (mJy)} & \colhead{Period (s)} & \colhead{$DM$ (cm$^{-3}$pc)} & \colhead{Distance (kpc)}
}
\startdata
\cutinhead{Pulsar matches with the sample of outliers}

    10 24 38.698  -07 19 19.07 & 0.17 &   J1024-0719 &   5.21 &  0.005 &  6.49 & 0.53 \\
    09 22 14.008  +06 38 22.84 & 0.51 &   J0922+0638 &  10.33 &  0.431 & 27.27 & 1.20 \\
    10 22 58.011  +10 01 52.85 & 0.40 &   J1022+1001 &   3.69 &  0.016 & 10.25 & 0.40 \\
    12 39 40.386  +24 53 49.87 & 1.19 &   J1239+2453 &  11.53 &  1.382 &  9.24 & 0.86 \\
    08 26 51.438  +26 37 22.83 & 1.19 &   J0826+2637 &  11.14 &  0.531 & 19.45 & 0.36 \\
    16 52 03.080  +26 51 39.85 & 0.56 &   J1652+2651 &   6.27 &  0.916 & 40.80 & 2.93 \\
    15 18 16.831  +49 04 34.19 & 0.31 &   J1518+4904 &   5.03 &  0.041 & 11.61 & 0.70 \\
    10 12 33.387  +53 07 02.09 & 0.66 &   J1012+5307 &   2.20 &  0.005 &  9.02 & 0.52 \\
    15 09 25.675  +55 31 32.90 & 0.62 &   J1509+5531 &  10.02 &  0.740 & 19.61 & 2.13 \\

\cutinhead{Pulsar matches passing variability criteria but not strictly point-like}

    16 07 12.078  -00 32 40.98 & 0.42 &   J1607-0032 &   3.93 &  0.422 & 10.68 & 0.59 \\
    10 23 47.622  +00 38 41.60 & 1.09 &   J1023+0038 &   3.12 &  0.002 & 14.32 & 0.90 \\
    09 43 30.092  +16 31 34.67 & 2.32 &   J0943+1631 &   1.51 &  1.087 & 20.32 & 1.76 \\
    16 40 16.699  +22 24 08.98 & 0.60 &   J1640+2224 &   1.92 &  0.003 & 18.43 & 1.19 \\

\cutinhead{Other pulsar matches in the {\sl FIRST} survey}

    09 53 09.287  +07 55 35.94 & 0.38 &   J0953+0755 &  83.22 &  0.253 &  2.96 & 0.26 \\
    15 43 38.837  +09 29 16.52 & 0.26 &   J1543+0929 &   6.17 &  0.748 & 35.24 & 7.69 \\
    07 51 09.148  +18 07 38.73 & 0.17 &   J0751+1807 &   1.42 &  0.003 & 30.25 & 0.62 \\
    11 15 38.456  +50 30 12.68 & 0.67 &   J1115+5030 &   1.00 &  1.656 &  9.20 & 0.54 \\

\enddata
\tablenotetext{a}{Data obtained from the ATNF Pulsar Catalog}
\end{deluxetable}

\end{document}